\documentclass[11pt]{article}
\usepackage{jheppub}
\usepackage{amsmath,amssymb,amsfonts}
\usepackage{mathalpha}
\usepackage{graphicx} % Required for inserting images
\usepackage{xcolor}
\usepackage{multirow}
\usepackage{tikz}
\usepackage{tikz-feynman}
\usetikzlibrary{tikzmark,calc}
\usetikzlibrary{decorations.pathmorphing,positioning}
\usepackage{subfigure}
\usepackage{mlx}
\usepackage[normalem]{ulem}
\usepackage{comment}
\tikzset{
    photon/.style={decorate, decoration={snake}},
    charged/.style={postaction={decorate},
        decoration={markings,mark=at position .55 with {\arrow[draw]{stealth}}}},
    gluon/.style={decorate, draw=black,
        decoration={coil,amplitude=4pt, segment length=5pt}}
}

\def\bea{\begin{eqnarray}}
\def\eea{\end{eqnarray}}
\def\<{\langle}
\def\>{\rangle}
\def\tr{\text{tr}}

\def\be{\begin{equation}}
\def\ee{\end{equation}}

\title{Gauged Soft Recursion: On-Shell Construction of Goldstone-Gauge Amplitudes}

\author[b,c]{Ian Low,}
\author[a]{Ming-Lei Xiao,}
\author[d]{and Yu-Hui Zheng}
\affiliation[a]{School of Science, Sun Yat-sen University,
Shenzhen 518100, China}
\affiliation[b]{High Energy Physics Division, Argonne National Laboratory,
Argonne, IL 60439, U.S.A.}
\affiliation[c]{Department of Physics and Astronomy, Northwestern University,
Evanston, IL60208, USA}
\affiliation[d]{School of Physics, Korea Institute for Advanced Study (KIAS), 85 Hoegi-ro, Seoul 02455, Korea}

\emailAdd{ilow@northwestern.edu}
\emailAdd{xiaomlei@mail.sysu.edu.cn}
\emailAdd{zhengyuhui@kias.re.kr}

\abstract{We present a new on-shell recursion relation for scattering amplitudes involving Nambu-Goldstone bosons with a gauged unbroken symmetry. 
A central challenge is that gauge interactions break Adler's zero condition for charged scalars, invalidating the standard soft recursion \cite{Cheung:2015ota}.
To overcome this, we introduce a ``gauged soft recursion'' that leverages the soft theorems of the gauge bosons themselves, combined with a novel decomposition of amplitudes into gauge-invariant components where Adler's zero is partially restored. The formalism, which also incorporates internal gauge bosons via angular momentum constraints, enables the systematic construction of tree-level amplitudes with arbitrary numbers of Goldstone bosons and gauge bosons in both Abelian and non-Abelian theories, as we demonstrate with explicit examples.
}

\begin{document}

\maketitle

\section{Introduction}

The study of on-shell recursion relations for scattering amplitudes is driven by the need for more efficient computational techniques in quantum field theory. Traditional Feynman diagram methods become unwieldy as the number of particles increases, leading to complex calculations with numerous terms. Recursion relations, such as the Britto-Cachazo-Feng-Witten (BCFW) approach \cite{Britto:2005fq,Britto:2004ap}, provide a streamlined method by breaking down complex amplitudes into simpler components derived from lower-point amplitudes. These developments have not only led to significant computational advancements but have also deepened our theoretical understanding of the underlying symmetries and structures in gauge theories \cite{Arkani-Hamed:2008bsc}. 
However, the applicability of these recursion relations is limited by the on-shell constructibility, which comes from the requirement that the amplitude vanishes at complex infinity of the momentum shift used in the recursion. Typically, effective field theories with higher derivative couplings are not constructible \cite{Cohen:2010mi,Cheung:2015cba}. One way to understand it is that new independent parameters are introduced for these effective operators that cannot be distinguished from the pole contributions constructed from on-shell unitarity. There are, however, cases where these parameters are constrained by symmetries, for example the shift symmetry of the  scalars, which give rise to vanishing amplitudes when their momenta are taken soft \cite{Adler:1964um,Cheung:2014dqa, Low:2014nga}. The recursion relations for these soft scalar theories, with various soft degrees, are constructed systematically in \cite{Cheung:2015ota,Luo:2015tat,Low:2019ynd,Cheung:2021yog}, which introduce rescaling factors to suppress the large circle contributions. 

In the real world, the most prominent example of scalars with vanishing soft behaviors is the pions in low-energy QCD. These are pseudo-Nambu-Goldstone bosons and would be exactly massless in the chiral limit. The effective theory describing the interactions of pions and baryons is the Chiral Perturbation Theory (ChPT) \cite{Weinberg:1968de,Weinberg:1978kz}, which is a nonlinear sigma model (NLSM) based on the symmetry breaking pattern $SU(2)_L\times SU(2)_R\to SU(2)_V$. The pions transform as the adjoint representation under the unbroken group $H=SU(2)_V$. Indeed, soft recursion relations have been applied to the pure mesonic sector of ChPT to very high orders in the derivative expansions \cite{Dai:2020cpk,Low:2022iim,Song:2024fae,Li:2024ghg}. 
In addition, nonlinearly realized symmetries have also played an important role in understanding the naturalness issue in the lightness of the Higgs boson. In this scenario, the 125 GeV Higgs boson arises as a pseudo-Nambu-Goldstone boson \cite{Kaplan:1983fs,Kaplan:1983sm,ArkaniHamed:2001nc,ArkaniHamed:2002qx,ArkaniHamed:2002qy,Contino:2003ve,Agashe:2004rs} and its interaction is described by a nonlinear effective field theory \cite{Giudice:2007fh}. More generally, techniques and insights from the scattering amplitude community have uncovered new and hidden structures in NLSMs  describing interactions of Nambu-Goldstone bosons \cite{Low:2015ogb,Low:2017mlh,Low:2018acv,Bijnens:2019eze,Low:2019wuv,Kampf:2019mcd,Kampf:2021jvf,Rodina:2021isd,Bartsch:2022pyi,Bartsch:2024ofb,Li:2024ghg,Sun:2022ssa,Sun:2022snw}.

However, the existing soft recursion relations have so far be limited to the neutral scalar sector. When gauge fields are involved, as when pions in ChPT carry electric charge and interact with photons, the presence of the gauge field breaks the shift symmetry, violating the Adler's zero condition for charged scalars and preventing the application of standard recursion techniques.
Nevertheless, the shift symmetry is broken in a specific manner -- through the replacement of ordinary derivatives with gauge-covariant derivatives -- while the nonlinear operator structure is otherwise preserved. 
This suggests that bootstrapping Goldstone amplitudes in the presence of gauge bosons should indeed be possible. 
In this work, we achieve this by extending the soft recursion relations through several novel techniques. 
First, we take into account the scaling factors for the gauge bosons at large complex momenta, whose massless poles are dictated by the soft photon theorems \cite{Weinberg:1964ew,Luo:2015tat}. 
Next we decompose amplitudes into gauge-invariant components $\mc{M} = \sum_i Q_i \mc{A}_i$ which can be individually constructed; the decomposition in the cases of multiple photons and non-Abelian gauge groups are also discussed. 
In each component, the external charged scalars not interacting with the gauge bosons behave effectively as neutral scalars and the Adler's zero condition is restored. In this way, they can be constructed in the extended soft recursion relation as in eq.~\eqref{eq:master}
\eq{
\mc{A}_{i} = - \sum_I\res_{z=z_I^\pm}\frac{\hat{\mc{A}}_{I\rm{L}}(z)\times\hat{\mc{A}}_{I\rm{R}}(z)}{zF(z)\hat{P}(z)^2} - \sum_s\res_{z=\frac{1}{a_{s}}} \frac{\hat{\mathcal{A}}_{i}(z)}{zF(z)}\ ,
}
which contain contributions both from the hard poles $z=z_I^\pm$ as appeared in the original soft recursion relation, and from the soft poles $z=1/a_s$ for each of the gauge bosons.
Finally, the inevitable photon exchange contributions can be included by demanding the total angular momentum $J=1$ for the residues at the photon poles.

This paper is organized as follows: In section 2, we review the standard soft recursion relation for pure Goldstone boson amplitudes in the NLSMs and ChPT. In section 3, we generalize this recursion to incorporate gauge interactions, introducing the soft photon theorem, the principle of charge decomposition, and the treatment of internal gauge bosons. In section 4, we present the gauged soft recursion relations for amplitudes with one or more photons, demonstrating their validity across different effective field theories. Finally, we summarize our results and discuss future directions.

\section{The Soft Recursion for Nambu-Goldstone Bosons}
\label{sec:soft_RR}

% general overview of recursion relations
By deforming external momenta by a complex variable $z$, BCFW recursion expresses an $n$-point amplitude in terms of lower-point amplitudes, significantly simplifying calculations in Yang-Mills theory and gravity \cite{Britto:2004ap,Britto:2005fq}. 
Despite its success, the applicability of on-shell recursion is constrained by the constructibility condition — namely, the requirement that amplitudes vanish as $z\to\infty$ for the chosen momentum deformation. But the constructibility conditions generally fail in effective field theories (EFTs) due to the presence of higher-derivative interactions and contact terms, which introduce additional independent parameters that cannot be fully reconstructed from factorization properties \cite{Cohen:2010mi,Cheung:2015cba,Cheung:2016drk}. 
Nevertheless, certain EFTs, such as ChPT, still admit modified recursion relations by leveraging the soft limit properties of Goldstone bosons \cite{Cheung:2014dqa,Cheung:2015ota}. In particular, the Adler’s zero condition states that in the exact chiral limit, the amplitude for a single soft Goldstone boson must vanish \cite{Adler:1964um}:
\begin{align}
    \mathcal{M}(\pi(p),\dots)\stackrel{p\to 0}{\longrightarrow}0.
\end{align}
This condition can be generalized by specifying the power $\sigma$ of the soft momentum in the limit $\mathcal{M}\sim p^{\sigma}$, where $\sigma$ is a positive integer known as the soft degree. This describes special models like the Dirac-Born-Infeld (DBI) and special Galileon (sGal) theories.

% soft block
The Goldstone bosons come from the coset of a symmetry breaking pattern $G/H$, and form a multiplet $\phi^i$ under some representation of the unbroken group $H$. We will use the notation $s_{ij\dots}=(p_i + p_j + \dots)^2$ throughout the paper. The leading order 4-point amplitude is then given by,
\begin{align}\label{eq:phi4}
    \mathcal{M}(\phi^{i_1},\phi^{i_2},\phi^{i_3},\phi^{i_4})= \frac{1}{f^2}\left(C^{i_1i_2i_3i_4}s_{12}+\text{permutations}\right) + O(p^4) \ .
\end{align}
The coefficient $C$ is a flavor tensor, invariant under the transformation of group $H$. The on-shell amplitude satisfies the Adler's zero condition for all four external momenta:
\begin{equation}
    \lim_{p_i\to0}s_{12} = \lim_{p_i\to0}s_{34} = 0 \ ,\quad i=1,2,3,4\ .
\end{equation}
They are known as soft blocks \cite{Low:2019ynd,Low:2022iim} and serve as the starting points of soft recursion relations. 

% soft recursion
Given the 4-point amplitude $\mc{M}_4$, we briefly review the on-shell construction of the $n$-point tree-level amplitudes $\mc{M}_n\equiv\mc{M}^{\rm tree}(\phi_1,\phi_2,\dots,\phi_n)$ by the soft recursion relation introduced in Ref.~\cite{Cheung:2015ota}. 
To establish the soft recursion, we begin with an all-line shift of the external momenta and define a soft factor $F_n(z)$ as:
\begin{align}\label{eq:softshift}
    \hat{p}_i(z)=(1-a_iz)p_i,\quad  i=1,\cdots,n \ , \\
    F_n(z)=\prod_{i=1}^n(1-a_iz)^{\sigma_i},\quad
\end{align}
where $\sigma_i$ is the soft degree of the scalar $\phi_i$. The constants $a_i$'s are determined by imposing momentum conservation:
\begin{align}\label{eq:p_conserve}
    \sum_i \hat{p}^\mu_i = \sum_i p^\mu_i =0 \ , \quad \Rightarrow \quad \sum_i a_i p^\mu_i = 0 \ .
\end{align}
In $D$-dimensional spacetime, the above equation can be viewed as a set of $D$ linear equations in $n$ variables $a_i$. When $n\le D$, the equations are undetermined for generic momenta; if $n=D+1$, the only solution is that all $a_i$ are equal, which is trivial. A non-trivial solution exists only when $n > D+1$, thus in $D=4$ we need $n\ge 6$ \cite{Low:2019ynd}. The self-consistency of soft recursion relations ensures the resulting amplitude is independent of the choice of the solution $\{a_i\}$.
Under this momentum shift, the amplitude is analytically continued to  $\hat{\mc{M}}(z)\equiv \mc{M}(\{\hat{p}_i(z)\})$ with two known properties: 1). It may have poles at the solutions $z_I^\pm$ of the quadratic equations
\begin{align}
    \hat{P}_I(z)^2=(P_I+Q_Iz)^2=0,
\end{align}
where $P_I = \sum_{i\in I}p_i$ is the total momentum at a certain scattering channel $I$\footnote{The channel is specified by the collection of particle labels $I$ on one of its two sides. To avoid redundancy, it can be chosen to be the side containing particle 1.} and $Q_I = \sum_{i\in I}a_ip_i$ is its shift;
2). It vanishes as $(z-1/a_i)^{\sigma_i}$ near $z=1/a_i$ for soft degree $\sigma_i$, which we take as $\sigma_i=1$ for all the Goldstone bosons.
Therefore, applying Cauchy's theorem we obtain: 
\begin{align}\begin{split}
    \mathcal{M}_n &=\frac{1}{2\pi \i}\oint_{z=0}\frac{dz}{z}\frac{\hat{\mathcal{M}}_n(z)}{F_n(z)}\\
    &= -\left(\sum_I\res_{z=z_I^{\pm}} + \sum_i\res_{z=1/a_i}\right)\frac{\hat{\mathcal{M}}(z)}{zF_n(z)} + B_\infty \ .
\end{split}\end{align}
If the shifted amplitude at $z=\infty$ is sufficiently suppressed by the soft factor $F_n(z)$ in the denominator, which does not introduce new poles due to the soft behaviors of the scalars,
\eq{
    B_\infty \sim \lim_{z\to \infty}\frac{\hat{\mathcal{M}}(z)}{F_n(z)} = 0 \ ,\qquad \lim_{z\to 1/a_i}\frac{\hat{\mathcal{M}}(z)}{F_n(z)} = \text{finite} \ ,
}
the amplitude would only depend on the poles of the propagators $z_I^\pm$, which can be on-shell constructed by the principle of unitarity:
\begin{align}\begin{split}\label{eq:phi6SR}
    \mathcal{M}_n &= -\sum_I \res_{z=z_I^{\pm}} \frac{\hat{\mathcal{M}}(z)}{zF_n(z)} 
    = -\sum_I \res_{z=z_I^{\pm}}\frac{\sum_i\hat{\mathcal{M}}^{I,i}(z)\hat{\mathcal{M}}^{\bar{I},i}(z)}{z\hat{P}_I^2(z)F_n(z)}\ .
\end{split}\end{align}
In the factorization, the sum over intermediate flavors $i$ contracts the flavor tensors in both subamplitudes. To illustrate this, we first demonstrate with an example that computes amplitude among specific flavor components, and then show how flavor-ordered amplitudes can be constructed recursively in a vastly simplified way.

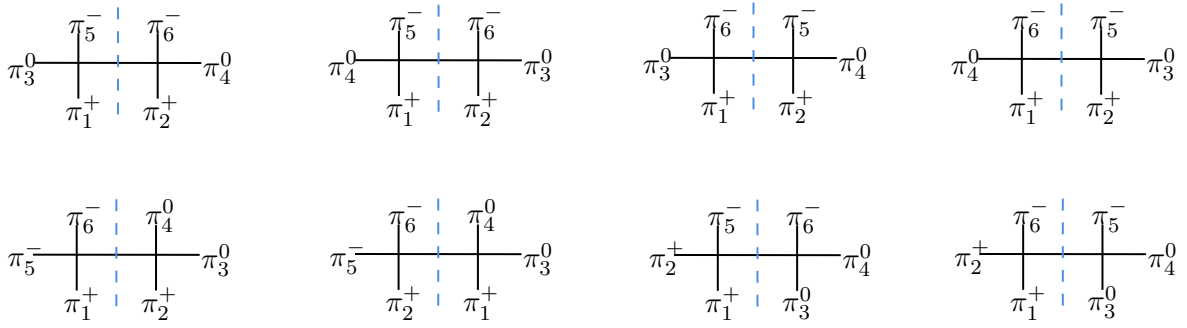
\begin{figure}[tbp]
    \centering

\tikzset{every picture/.style={line width=0.75pt}} %set default line width to 0.75pt        

\begin{tikzpicture}[x=0.75pt,y=0.75pt,yscale=-1,xscale=1]
%uncomment if require: \path (0,197); %set diagram left start at 0, and has height of 197

%Straight Lines [id:da09856712898268194] 
\draw    (34,47.6) -- (118,47.7) ;
%Straight Lines [id:da9560885110609548] 
\draw    (56,32.7) -- (56,65.7) ;
%Straight Lines [id:da8724762635932447] 
\draw    (96,32.7) -- (96,65.7) ;
%Straight Lines [id:da7760877219048963] 
\draw [color={rgb, 255:red, 74; green, 144; blue, 226 }  ,draw opacity=1 ] [dash pattern={on 4.5pt off 4.5pt}]  (76,19.65) -- (76,75.65) ;
%Straight Lines [id:da5807070367373044] 
\draw    (195.67,46.27) -- (279.67,46.37) ;
%Straight Lines [id:da17396869074120436] 
\draw    (217.67,31.37) -- (217.67,64.37) ;
%Straight Lines [id:da7813187754176082] 
\draw    (257.67,31.37) -- (257.67,64.37) ;
%Straight Lines [id:da020821681585709295] 
\draw [color={rgb, 255:red, 74; green, 144; blue, 226 }  ,draw opacity=1 ] [dash pattern={on 4.5pt off 4.5pt}]  (237.67,18.32) -- (237.67,74.32) ;
%Straight Lines [id:da45351533506762687] 
\draw    (354.5,45.1) -- (438.5,45.2) ;
%Straight Lines [id:da357452835064489] 
\draw    (376.5,30.2) -- (376.5,63.2) ;
%Straight Lines [id:da5236564713683152] 
\draw    (416.5,30.2) -- (416.5,63.2) ;
%Straight Lines [id:da522861157679766] 
\draw [color={rgb, 255:red, 74; green, 144; blue, 226 }  ,draw opacity=1 ] [dash pattern={on 4.5pt off 4.5pt}]  (396.5,17.15) -- (396.5,73.15) ;
%Straight Lines [id:da10841150693901214] 
\draw    (509.83,45.6) -- (593.83,45.7) ;
%Straight Lines [id:da7993182333918486] 
\draw    (531.83,30.7) -- (531.83,63.7) ;
%Straight Lines [id:da8284078618925391] 
\draw    (571.83,30.7) -- (571.83,63.7) ;
%Straight Lines [id:da284690116264977] 
\draw [color={rgb, 255:red, 74; green, 144; blue, 226 }  ,draw opacity=1 ] [dash pattern={on 4.5pt off 4.5pt}]  (551.83,17.65) -- (551.83,73.65) ;
%Straight Lines [id:da3164538133288475] 
\draw    (33.17,144.43) -- (117.17,144.53) ;
%Straight Lines [id:da8702935811136909] 
\draw    (55.17,129.53) -- (55.17,162.53) ;
%Straight Lines [id:da5566936339923536] 
\draw    (95.17,129.53) -- (95.17,162.53) ;
%Straight Lines [id:da025504410785291087] 
\draw [color={rgb, 255:red, 74; green, 144; blue, 226 }  ,draw opacity=1 ] [dash pattern={on 4.5pt off 4.5pt}]  (75.17,116.48) -- (75.17,172.48) ;
%Straight Lines [id:da8299754732553517] 
\draw    (195.5,144.1) -- (279.5,144.2) ;
%Straight Lines [id:da31786965392181343] 
\draw    (217.5,129.2) -- (217.5,162.2) ;
%Straight Lines [id:da35091098181464864] 
\draw    (257.5,129.2) -- (257.5,162.2) ;
%Straight Lines [id:da8615955560849158] 
\draw [color={rgb, 255:red, 74; green, 144; blue, 226 }  ,draw opacity=1 ] [dash pattern={on 4.5pt off 4.5pt}]  (237.5,116.15) -- (237.5,172.15) ;
%Straight Lines [id:da5708431360962087] 
\draw    (356.5,144.6) -- (440.5,144.7) ;
%Straight Lines [id:da07901403972217047] 
\draw    (378.5,129.7) -- (378.5,162.7) ;
%Straight Lines [id:da814596770535389] 
\draw    (418.5,129.7) -- (418.5,162.7) ;
%Straight Lines [id:da5634970239107994] 
\draw [color={rgb, 255:red, 74; green, 144; blue, 226 }  ,draw opacity=1 ] [dash pattern={on 4.5pt off 4.5pt}]  (398.5,116.65) -- (398.5,172.65) ;
%Straight Lines [id:da7542076829922071] 
\draw    (510.5,144.1) -- (594.5,144.2) ;
%Straight Lines [id:da5942072783214559] 
\draw    (532.5,129.2) -- (532.5,162.2) ;
%Straight Lines [id:da1784083462735545] 
\draw    (572.5,129.2) -- (572.5,162.2) ;
%Straight Lines [id:da6263358962677854] 
\draw [color={rgb, 255:red, 74; green, 144; blue, 226 }  ,draw opacity=1 ] [dash pattern={on 4.5pt off 4.5pt}]  (552.5,116.15) -- (552.5,172.15) ;

% Text Node
\draw (48,62.4) node [anchor=north west][inner sep=0.75pt]    {$\pi _{1}^{+}$};
% Text Node
\draw (87,62.4) node [anchor=north west][inner sep=0.75pt]    {$\pi _{2}^{+}$};
% Text Node
\draw (18.67,41.4) node [anchor=north west][inner sep=0.75pt]    {$\pi _{3}^{0}$};
% Text Node
\draw (117.33,40.73) node [anchor=north west][inner sep=0.75pt]    {$\pi _{4}^{0}$};
% Text Node
\draw (49.33,18.4) node [anchor=north west][inner sep=0.75pt]    {$\pi _{5}^{-}$};
% Text Node
\draw (89,18.4) node [anchor=north west][inner sep=0.75pt]    {$\pi _{6}^{-}$};
% Text Node
\draw (209.67,61.07) node [anchor=north west][inner sep=0.75pt]    {$\pi _{1}^{+}$};
% Text Node
\draw (248.67,61.07) node [anchor=north west][inner sep=0.75pt]    {$\pi _{2}^{+}$};
% Text Node
\draw (180.33,40.07) node [anchor=north west][inner sep=0.75pt]    {$\pi _{4}^{0}$};
% Text Node
\draw (279,39.4) node [anchor=north west][inner sep=0.75pt]    {$\pi _{3}^{0}$};
% Text Node
\draw (211,17.07) node [anchor=north west][inner sep=0.75pt]    {$\pi _{5}^{-}$};
% Text Node
\draw (250.67,17.07) node [anchor=north west][inner sep=0.75pt]    {$\pi _{6}^{-}$};
% Text Node
\draw (368.5,59.9) node [anchor=north west][inner sep=0.75pt]    {$\pi _{1}^{+}$};
% Text Node
\draw (407.5,59.9) node [anchor=north west][inner sep=0.75pt]    {$\pi _{2}^{+}$};
% Text Node
\draw (339.17,38.9) node [anchor=north west][inner sep=0.75pt]    {$\pi _{3}^{0}$};
% Text Node
\draw (437.83,38.23) node [anchor=north west][inner sep=0.75pt]    {$\pi _{4}^{0}$};
% Text Node
\draw (369.83,15.9) node [anchor=north west][inner sep=0.75pt]    {$\pi _{6}^{-}$};
% Text Node
\draw (409.5,15.9) node [anchor=north west][inner sep=0.75pt]    {$\pi _{5}^{-}$};
% Text Node
\draw (523.83,60.4) node [anchor=north west][inner sep=0.75pt]    {$\pi _{1}^{+}$};
% Text Node
\draw (562.83,60.4) node [anchor=north west][inner sep=0.75pt]    {$\pi _{2}^{+}$};
% Text Node
\draw (494.5,39.4) node [anchor=north west][inner sep=0.75pt]    {$\pi _{4}^{0}$};
% Text Node
\draw (593.17,38.73) node [anchor=north west][inner sep=0.75pt]    {$\pi _{3}^{0}$};
% Text Node
\draw (525.17,16.4) node [anchor=north west][inner sep=0.75pt]    {$\pi _{6}^{-}$};
% Text Node
\draw (564.83,16.4) node [anchor=north west][inner sep=0.75pt]    {$\pi _{5}^{-}$};
% Text Node
\draw (47.17,159.23) node [anchor=north west][inner sep=0.75pt]    {$\pi _{1}^{+}$};
% Text Node
\draw (86.17,159.23) node [anchor=north west][inner sep=0.75pt]    {$\pi _{2}^{+}$};
% Text Node
\draw (18.83,136.23) node [anchor=north west][inner sep=0.75pt]    {$\pi _{5}^{-}$};
% Text Node
\draw (116.5,137.57) node [anchor=north west][inner sep=0.75pt]    {$\pi _{3}^{0}$};
% Text Node
\draw (48.5,115.23) node [anchor=north west][inner sep=0.75pt]    {$\pi _{6}^{-}$};
% Text Node
\draw (88.17,115.23) node [anchor=north west][inner sep=0.75pt]    {$\pi _{4}^{0}$};
% Text Node
\draw (209.5,158.9) node [anchor=north west][inner sep=0.75pt]    {$\pi _{2}^{+}$};
% Text Node
\draw (248.5,158.9) node [anchor=north west][inner sep=0.75pt]    {$\pi _{1}^{+}$};
% Text Node
\draw (181.17,135.9) node [anchor=north west][inner sep=0.75pt]    {$\pi _{5}^{-}$};
% Text Node
\draw (278.83,137.23) node [anchor=north west][inner sep=0.75pt]    {$\pi _{3}^{0}$};
% Text Node
\draw (210.83,114.9) node [anchor=north west][inner sep=0.75pt]    {$\pi _{6}^{-}$};
% Text Node
\draw (250.5,114.9) node [anchor=north west][inner sep=0.75pt]    {$\pi _{4}^{0}$};
% Text Node
\draw (370.5,159.4) node [anchor=north west][inner sep=0.75pt]    {$\pi _{1}^{+}$};
% Text Node
\draw (409.5,159.4) node [anchor=north west][inner sep=0.75pt]    {$\pi _{3}^{0}$};
% Text Node
\draw (342.17,136.4) node [anchor=north west][inner sep=0.75pt]    {$\pi _{2}^{+}$};
% Text Node
\draw (439.83,137.73) node [anchor=north west][inner sep=0.75pt]    {$\pi _{4}^{0}$};
% Text Node
\draw (371.83,115.4) node [anchor=north west][inner sep=0.75pt]    {$\pi _{5}^{-}$};
% Text Node
\draw (411.5,115.4) node [anchor=north west][inner sep=0.75pt]    {$\pi _{6}^{-}$};
% Text Node
\draw (524.5,158.9) node [anchor=north west][inner sep=0.75pt]    {$\pi _{1}^{+}$};
% Text Node
\draw (563.5,158.9) node [anchor=north west][inner sep=0.75pt]    {$\pi _{3}^{0}$};
% Text Node
\draw (496.17,135.9) node [anchor=north west][inner sep=0.75pt]    {$\pi _{2}^{+}$};
% Text Node
\draw (593.83,137.23) node [anchor=north west][inner sep=0.75pt]    {$\pi _{4}^{0}$};
% Text Node
\draw (525.83,114.9) node [anchor=north west][inner sep=0.75pt]    {$\pi _{6}^{-}$};
% Text Node
\draw (565.5,114.9) node [anchor=north west][inner sep=0.75pt]    {$\pi _{5}^{-}$};

\end{tikzpicture}

    \caption{Here are all possible channels for the recursion of the tree-level amplitude $\mathcal{M}(\pi_1^+,\pi_2^+,\pi_3^0,\pi_4^0,\pi_5^-,\pi_6^-)$. 
    }
    \label{fig:pi60}
\end{figure}

We now illustrate this with an amplitude among 6 specific flavored Goldstone bosons $\mc{M}_6(\pi_1^+,\pi_2^+,\pi_3^0,\pi_4^0,\pi_5^-,\pi_6^-)$. Here, $\pi^\pm=\pi^1\pm\i \pi^2$ and $\pi^0$ form a triplet of $SU(2) \subset H$ and are labelled by their charges under a $U(1)$ subgroup\footnote{The charge is defined for future reference, as we will be gauging the $U(1)$ later.}. The recursion formula in eq.~\eqref{eq:phi6SR} gives:
\eq{ 
\mc{M}_6 = -\sum_I\res_{z=z^{\pm}_I}\frac{\hat{\mc{M}}^I_4(z)\times\hat{\mc{M}}^{\bar{I}}_4(z)}{z \hat{P}_I^2(z)F_6(z)} \ .
}
The unique 4-point soft block is $\mc{M}_4(\pi_1^{i_1},\pi_2^{i_2},\pi_3^{i_3},\pi_4^{i_4})=\delta^{i_1i_2}\delta^{i_3i_4}s_{12}/f^2+\text{permutations}$, with the following relevant flavor components
\eq{\label{eq:pi4_bb}
\mc{M}_4(\pi^+_1,\pi^-_2,\pi^0_3,\pi^0_4)=\frac{1}{f^2}s_{12}\ ,\quad \mc{M}_4(\pi^+_1,\pi^+_2,\pi^-_3,\pi^-_4)=-\frac{1}{f^2}s_{12}\ .
}
The available building blocks restrict the allowed factorization channels. For instance, the channel $I=(123)$ is excluded because it requires a subamplitude $\mc{M}_4(\pi_1^+,\pi_2^+,\pi_3^0,\pi_4^i)$ that is a vanishing component in the flavor tensor. 
As a result, we are left with the 8 factorization channels shown in figure~\ref{fig:pi60}.
Since the subamplitudes are regular functions of $z$, we can apply Cauchy's theorem once more for each channel, evaluating the integral by taking residues at $z=0$ and the soft poles $z=1/a_i$. This yields:
\eq{\label{eq:M6RR}
    \mc{M}_6 &= \sum_I\left( \res_{z=0} + \sum_{i=1}^6\res_{z=1/a_i} \right) \frac{\hat{\mc{M}}^I_4(z)\times\hat{\mc{M}}^{\bar{I}}_4(z)}{z \hat{P}_I^2(z)F_6(z)} \\
    &= \sum_I \frac{\mc{M}^I_4\times\mc{M}^{\bar{I}}_4}{P_I^2} 
    + \sum_{i=1}^6 \res_{z=1/a_i} \frac{1}{zF_6(z)}\sum_I\frac{\hat{\mc{M}}^I_4(z)\times\hat{\mc{M}}^{\bar{I}}_4(z)}{ \hat{P}_I^2(z)} \\
    &\equiv \frac{1}{f^4}(\mc{M}_6^{(1)} + \mc{M}_6^{(2)}) \ .
}
The first term $\mc{M}_6^{(1)}$ corresponds to the pole diagrams from the eight factorization channels shown in figure~\ref{fig:pi60},
\eq{\label{eq:M6RR1}
    \mc{M}_6^{(1)} &\equiv f^4\sum_I \frac{\mc{M}^I_4\times\mc{M}^{\bar{I}}_4}{P_I^2} \\
    &= \frac{s_{15}s_{26}}{s_{135}}+\frac{s_{15}s_{26}}{s_{145}}+\frac{s_{16}s_{25}}{s_{136}}+\frac{s_{16}s_{25}}{s_{146}}-\frac{s_{34}s_{56}}{s_{156}}-\frac{s_{34}s_{56}}{s_{256}}-\frac{s_{12}s_{34}}{s_{125}}-\frac{s_{12}s_{34}}{s_{126}}\ ,
}
The second term $\mc{M}_6^{(2)}$ reproduces the 6-point contact interactions required by the nonlinearly realized shift symmetry \cite{Low:2014nga} as
\eq{\label{eq:M6RR2}
    \mc{M}_6^{(2)} &\equiv \sum_{i=1}^6 \res_{z=1/a_i} \frac{1}{zF_6(z)}
    \times f^4\sum_I\frac{\hat{\mc{M}}^I_4(z)\times\hat{\mc{M}}^{\bar{I}}_4(z)}{ \hat{P}_I^2(z)} \\
    &= \sum_{i=1}^6\res_{z=1/a_i}\frac{1}{zF_6(z)}\left(-\hat{s}_{12}-\hat{s}_{34}-\hat{s}_{56}\right) \\
    &= s_{12} + s_{34} + s_{56} \ .
}
This contribution precisely ensures that the full amplitude satisfies the Adler's zero condition for each external leg.

In ChPT with spontaneous symmetry breaking $SU(N)_L\times SU(N)_R \to SU(N)$, the recursion relations are particularly powerful when formulated in terms of flavor-ordered partial amplitudes. The Goldstone bosons constitute the adjoint representation of the unbroken group $H=SU(N)$ as $\phi = \sum_A\phi^A T^A$, with $T^A$ the generators of $SU(N)$. 
It is convenient to define the 6-point flavor-ordered partial amplitude $\mc{A}$: 
\eq{
    \mathcal{M}_6 = \tr\left(T^{A_1}T^{A_2}T^{A_3}T^{A_4}T^{A_5}T^{A_6}\right)\mc{A}(p_1,p_2,p_3,p_4,p_5,p_6) + \text{permutations}\ .
}
For the tensor contraction in eq.~\eqref{eq:phi6SR}, we may use the completeness condition \cite{Low:2019ynd}
\eq{
    & \sum_{B}\tr\left(T^{A_1}T^{A_2}T^{A_3}T^{B}\right) \times \tr\left(T^{B}T^{A_4}T^{A_5}T^{A_6}\right) = \\
    & \tr\left(T^{A_1}T^{A_2}T^{A_3}T^{A_4}T^{A_5}T^{A_6}\right) - \frac{1}{6}\tr\left(T^{A_1}T^{A_2}T^{A_3}\right)\times \tr\left(T^{A_4}T^{A_5}T^{A_6}\right)
}
where the second term corresponds to the $U(1)$ component of a $U(N)$ symmetry. This component decouples from the $SU(N)$ multiplet and does not contribute to the tree-level amplitude after summing over all the channels. Therefore, the partial amplitude can be factorized similar to the full amplitude, but involves only three factorization channels~\cite{Cheung:2015ota}
\eq{
    \mc{A} = \mc{A}^{(123)} + \mc{A}^{(126)} + \mc{A}^{(156)} \quad, \qquad \mc{A}^{I} = -\res_{z=z_I^{\pm}}\frac{\hat{\mathcal{A}}_{\rm L}^{I}(z)\hat{\mathcal{A}}_{\rm R}^{\bar{I}}(z)}{z\hat{P}_I^2(z)F_n(z)}\ ,
}
which is a significant simplification compared to the 8 channels in the previous calculation.
It has also been successfully applied to Goldstone bosons in other representations, such as the (anti-)fundamental of $H=SU(N)\times U(1)$ in the composite Higgs models \cite{Low:2014oga}.
The leading order 4-point amplitude of the complex scalar multiplet $\phi^i\in \mathbf{N}$ is
\eq{\label{eq:SUN4p}
    \mc{M}_4(\phi^{i_1},\phi^{i_2},\bar\phi_{i_3},\bar\phi_{i_4}) = \frac{1}{f^2}(\delta^{i_1}_{i_3}\delta^{i_2}_{i_4} + \delta^{i_1}_{i_4}\delta^{i_2}_{i_3})s_{12} \ ,
}
where $\bar\phi_i\in \bar{\mathbf{N}}$ is the charge conjugate of $\phi^i$ and carries opposite $U(1)$ charge as well.
In this case, the 6-point amplitude has the flavor structure \input{Fig1}
\begin{align}\begin{split}\label{eq:phi6gauge}
    \mathcal{M}_6(\phi^{i_1},\phi^{i_2},\phi^{i_3},\bar\phi_{i_4},\bar\phi_{i_5},\bar\phi_{i_6})=\delta^{i_1}_{i_4}\delta^{i_2}_{i_5}\delta^{i_3}_{i_6}\ \mc{A}+ \text{permutations of $(4,5,6)$}
\end{split}\end{align}
where the partial amplitude $\mc{A}$ can also be constructed using the soft recursion relation as
\eq{
    \mc{A} &= \mc{A}^{(124)} + \mc{A}^{(125)} + \mc{A}^{(134)} + \mc{A}^{(136)} + \mc{A}^{(145)} + \mc{A}^{(146)} \ , \\
    &\quad \mc{A}^{I} = -\res_{z=z_I^{\pm}}\frac{\hat{\mathcal{A}}_{\rm L}^{I}(z)\hat{\mathcal{A}}_{\rm R}^{\bar{I}}(z)}{z\hat{P}_I^2(z)F_n(z)}\ ,
}
in which the involved channels are depicted in figure~\ref{fig:phi6}. 
%Similar to eq.~\eqref{eq:M6RR}, 
The partial amplitude consists of two parts
\eq{\label{eq:SUN6p}
    \mc{A} &= \frac{1}{f^4}(\mc{A}^{(1)}+\mc{A}^{(2)}) \ ,\\
    \mc{A}^{(1)} &= \frac{s_{14}s_{36}}{s_{124}}+\frac{s_{25}s_{36}}{s_{125}}+\frac{s_{14}s_{25}}{s_{134}}+\frac{s_{25}s_{36}}{s_{136}}+\frac{s_{14}s_{36}}{s_{145}}+\frac{s_{14}s_{25}}{s_{146}} \ ,\\
    \mc{A}^{(2)} &= \sum_i \res_{z=\frac{1}{a_i}}\frac{1}{zF_6(z)}\left(
    \frac{\hat{s}_{14}\hat{s}_{36}}{\hat{s}_{124}}+\frac{\hat{s}_{25}\hat{s}_{36}}{\hat{s}_{125}}
    +\frac{\hat{s}_{14}\hat{s}_{25}}{\hat{s}_{134}}+\frac{\hat{s}_{25}\hat{s}_{36}}{\hat{s}_{136}}
    +\frac{\hat{s}_{14}\hat{s}_{36}}{\hat{s}_{145}}+\frac{\hat{s}_{14}\hat{s}_{25}}{\hat{s}_{146}}\right) \\
    &= -s_{14}-s_{25}-s_{36} \ ,
}
where again, in terms of the Lagrangian approach, $\mc{A}^{(1)}$ comes from the Feynman diagrams with poles, and $\mc{A}^{(2)}$ comes from the 6-point Feynman vertices predicted by the non-linear symmetry. The result is identical to the soft recursion relation for $SO(N)$ fundamentals obtained in \cite{Low:2019ynd}. We will come back to this example with the $U(1)$ being gauged and the Goldstone bosons $\phi^i$ form a charged multiplet.

\section{The Gauge Interaction of the Goldstone Bosons}

In both the Chiral Perturbation Theory and the composite Higgs model, the unbroken symmetry is (partially) gauged. This gauging promotes the scalars to pseudo-Goldstone bosons, as the gauge interaction explicitly breaks their shift symmetries. Consequently, in the scattering amplitude, the Adler's zero condition is lost for the charged scalars.
As a concrete example, consider the simplest nontrivial case of pions interacting with a photon, $\mc{M}_{4+1}(\pi^+,\pi^-,\pi^0,\pi^0,\gamma)$, where the $U(1)$ subgroup is gauged. 
Given the 4-point soft block in eq.~\eqref{eq:pi4_bb}, the Feynman rule calculation gives
\eq{\label{eq:5-pt_amp}
    \mc{M}_{4+1}(\pi^+,\pi^-,\pi^0,\pi^0,\gamma) = e(p_{5}^{\mu}\varepsilon^\nu - p_{5}^{\nu}\varepsilon^\mu)\frac{p_{1\mu} p_{2\nu}}{s_{15}s_{25}} \times \frac{s_{34}}{f^2}.
}
This amplitude satisfies the Adler's zero condition only for the neutral scalars $\pi^0$ (with momenta $p_{3,4}$), but not for the charged scalars ($p_{1,2}$). 
As a result, if we attempt to derive eq.~\eqref{eq:5-pt_amp} using on-shell recursion, we cannot include the soft factor $1-a_iz$ for the charged legs $i=1,2$ in $F(z)$ as in eq.~\eqref{eq:phi6SR}, because this would introduce unconstrained residues. In the extreme case of $\mc{M}_{4+1}(\pi^+,\pi^+,\pi^-,\pi^-,\gamma)$ where all the scalars are charged, there would be no available $F(z)$. Hence the gauge interaction introduces new challenges: they rob $F(z)$ of the necessary powers of $z$ to suppress the large-$z$ behavior, leading to a non-vanishing boundary term that obstructs the on-shell recursion. A naive conclusion would be that such theories are not on-shell constructible.

However, the theory remains highly constrained by the underlying shift symmetry, which is broken in a controlled manner solely by the gauge interaction. The non-linear symmetry still fixes the Wilson coefficients of higher-point operators in terms of the decay constant $f$. As we will demonstrate, a synthesis of the gauge boson soft theorems and the residual Adler's zero conditions provides the key to constructing recursion relations for amplitudes involving both Goldstone and gauge bosons.

\subsection{Soft Photon/Gluon Theorem}

Consider an $(n+1)$-particle on-shell amplitude $\mathcal{M}_{n+1}(\phi_1,\cdots \phi_n;\gamma(p_s))$ with an external gauge boson of momentum $p_s$ and helicity $h=+1$. Weinberg’s soft theorem states that in the soft limit $p_s\to \epsilon p_s$ with $\epsilon\to 0$, the amplitude behaves as \cite{Weinberg:1964ew}
\begin{align}
\label{eq:photonres}
    \mathcal{M}_{n+1}=\left(\frac{1}{\epsilon}S^{(0)}+S^{(1)}+O(\epsilon) \right)\mathcal{M}_n(\phi_1,\cdots \phi_n),
\end{align}
where $\mc{M}_n$ is the corresponding $n$-particle ``hard'' amplitude without the external photon. 
In the spinor-helicity formalism \cite{Elvang:2013cua}, the 4-momentum of the photon is denoted $(p_s)_{\alpha\dot\alpha}=|s\rangle_{\alpha}[s|_{\dot\alpha}$, and the soft limit is imposed by $|s\rangle\to \sqrt{\epsilon}|s\rangle $, $|s]\to \sqrt{\epsilon}|s]$. 
Under this limit, Weinberg’s theorem decomposes into a leading soft factor $S^{(0)}$ that scales like $\epsilon^{-1}$, and a subleading factor $S^{(1)}$ of order $\epsilon^0$. For a gauge boson of positive helicity, these soft factors take the form \cite{Cachazo:2014fwa,Elvang:2016qvq}:
\begin{align}
    S^{(0)}= e\sum_i \frac{\langle ri \rangle}{\langle rs\rangle\langle si\rangle}q_i,\quad 
    S^{(1)}= e\sum_i \frac{1}{\langle si\rangle}[s|_{\dot\alpha}\frac{\partial}{\partial [i|_{\dot\alpha}}q_i,
    \label{eq:soft_expansion}
\end{align}
where $q_i$ is the electric charge of the $i$-th particle. For simplicity, here we do not introduce any effective operators that can modify $S^{(1)}$. The auxiliary spinor $|r\rangle$ in $S^{(0)}$ is introduced that reflects the gauge dependence of the polarization vector. Due to gauge invariance, this dependence ultimately cancels out in full amplitudes, a process where the charge conservation $\sum_i q_i = 0$ plays a crucial role.
For the other helicity $h=-1$ of the gauge boson, one just swaps the angled brackets with the squared brackets in the formula.

We will also be using the non-Abelian version of the soft photon theorem, where the charge $q_i$ is replaced by the generator $T_i^a$, the $a$th generator matrix acting on the $i$th particle. The soft factors take the following form
\begin{align}
    S^{(0)}=g\sum_i \frac{\langle ri \rangle}{\langle rs\rangle\langle si\rangle}T^a_i \,,\quad 
    S^{(1)}=g\sum_i \frac{1}{\langle si\rangle}[s|_{\dot\alpha}\frac{\partial}{\partial [i|_{\dot\alpha}}T^a_i \,.
    \label{eq:soft_expansion_na}
\end{align}
A key difference in the non-Abelian case arises when multiple gauge bosons are present. The leading soft factor now includes a sum over all external particles, including the other gauge bosons (which transform under the adjoint representation with generators $(T^a_i)^{bc} = \i f^{abc}_i$. For example, the double-soft limit for two-gluons is given by:
\eq{
    \mc{M}_{n+2}(\phi_1,\dots,\phi_n;\gamma^a(\epsilon p_a),\gamma^b(p_b)) &= \frac{g}{\epsilon}\Big[\sum_{i=1}^n\frac{\vev{ri}}{\vev{ra}\vev{ai}}T^a_i\cdot\mc{M}_{n+1}(\phi_1,\dots,\phi_n;\gamma^b(p_b)) \\
    &\hspace{-20pt} + \frac{\vev{rb}}{\vev{ra}\vev{ab}}\sum_c(\i f^{abc})\mc{M}_{n+1}(\phi_1,\dots,\phi_n;\gamma^c(p_b))\Big] + O(\epsilon^0)
}

Under a momentum shift $p_s\to p_s(1-a_sz)$, the terms in the soft expansion scale as $S^{(\alpha)}\to (1-a_sz)^{\alpha-1}S^{(\alpha)}$. 
This scaling is central to incorporating gauge bosons into the soft recursion, as it dictates the form of the residue at the soft pole $z=1/a_s$:
\begin{equation}\begin{split}\label{eq:soft_photon_recursion}
    \res_{z=1/a_s}\frac{\hat{\mathcal{M}}_{n+1}(z)}{z(1-a_sz)^k}
    &= \res_{z=\frac{1}{a_s}}\left[\frac{1}{(1-a_s z)^{k+1}}\frac{S^{(0)}\hat{\mathcal{M}}_n(z)}{z} + \frac{1}{(1-a_s z)^{k}}\frac{S^{(1)}\hat{\mathcal{M}}_n(z)}{z} + \cdots\right]\ .
\end{split}\end{equation}
The series terminates at the term with $S^{(k)}$, beyond which there would be no residue at $z=1/a_s$.
Since there are universal formula only for $S^{(0)}$ and $S^{(1)}$ in the soft expansion, we can at most take $k=1$, including one power of $1-a_s z$ in the denominator for each gauge boson. Unlike the case of Goldstone bosons where the $F(z)$ does not introduce new poles, here the $z=1/a_s$ pole is a physical pole to begin with, and the extra factor in the denominator simply asks for the next-to-leading order soft behavior $S^{(1)}$ in the residue.

By dimensional counting, under the all-line momentum shift, if the $n$-Goldstone amplitude has $\lim_{z\to\infty}\hat{\mc{M}}_n \sim z^m$, adding a minimally coupled gauge boson would imply $\lim_{z\to\infty}$ $\hat{\mc{M}}_{n+1} \sim z^{m-1}$. In the denominator, we may have a factor of $1-a_i z$ for each of the neutral scalars (at most $n-2$ of them) and the gauge boson, which is at most a $(n-1)$-degree polynomial. Therefore, when the $n$-Goldstone amplitude $\mc{M}_n$ is on-shell constructible $m<n$, the $\mc{M}_{n+1}$ is also on-shell constructible if there is only a pair of charged scalars; the more charged scalars are involved, the less likely the amplitude can be obtained from the recursion relation. In order to construct the amplitudes for generically charged scalars, we need a new technique.

\subsection{The Principle of Amplitude Decomposition}
\label{sec:charge}

The previous analysis indicates that an amplitude $\mc{M}_{n+1}$ can be on-shell constructed when there is only a pair of charged scalars, but not necessarily when more are present. In this subsection, we show that while the latter may not be directly constructed, they can be systematically obtained from the former through a principled decomposition. 

To start, let's consider the four-scalar amplitudes in scalar QED, augmented by a quartic scalar potential $(\lambda/4!) \phi^4$, which involve two gauge invariant subamplitudes: the gauge interaction via photon exchange, which contains a factor  of $e^2$, and the  contact interaction multiplied by the quartic coupling  $\lambda$,
\begin{align}
    \mc{M}(\phi_1,\phi_2,\phi_3,\phi_4) = e^2\mc{A}_{e^2} + \lambda\mc{A}_\lambda \ .\label{eq:amp_scalar_QED} 
\end{align}
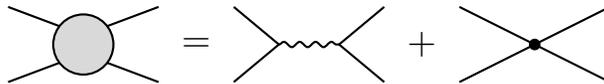
\[
\begin{tikzpicture}[baseline=(current bounding box.center)]
  % Blob diagram (left)
  \begin{scope}[scale=1]
    \draw[thick] (-1,0.5) -- (-0.2,0.2);    % Top-left leg
    \draw[thick] (-1,-0.5) -- (-0.2,-0.2);  % Bottom-left leg
    \draw[thick] (0.2,0.2) -- (1,0.5);      % Top-right leg
    \draw[thick] (0.2,-0.2) -- (1,-0.5);    % Bottom-right leg
    \filldraw[fill=gray!30, draw=black, thick] (0,0) circle (0.4); % Shaded blob
  \end{scope}
  
  % Equals sign
  \node at (1.5,0) {\Large$=$};
  
  % Photon exchange diagram (middle-right)
  \begin{scope}[shift={(3.0,0)}, scale=1]
    \draw[thick] (-1,0.5) -- (-0.4,0);    % Top-left leg
    \draw[thick] (-1,-0.5) -- (-0.4,0);  % Bottom-left leg
    \draw[thick] (0.4,0) -- (1,0.5);      % Top-right leg
    \draw[thick] (0.4,0) -- (1,-0.5);    % Bottom-right leg
    % Horizontal photon propagator (wavy line)
    \draw[decorate, decoration={snake, amplitude=1pt, segment length=6pt}, thick] 
      (-0.4,0) -- (0.4,0);
  \end{scope}
  
  % Plus sign
  \node at (4.5,0) {\Large$+$};
  
  % Contact diagram (right)
  \begin{scope}[shift={(6.0,0)}, scale=1]
    \draw[thick] (-1,0.5) -- (0,0);    % Top-left leg
    \draw[thick] (-1,-0.5) -- (0,0);   % Bottom-left leg
    \draw[thick] (0,0) -- (1,0.5);     % Top-right leg
    \draw[thick] (0,0) -- (1,-0.5);    % Bottom-right leg
    \filldraw[black] (0,0) circle (2pt); % Contact vertex
  \end{scope}
\end{tikzpicture}
\]
Since the two couplings $e$ and $\lambda$ are independent parameters, the two components are both physically valid amplitudes on their own. It is natural to compute each of them individually, and simply take their sum as the final result. 
Each component may be regarded as a physically valid amplitude of some particular model, where all the other parameters are set to zero.
For example, $\mc{A}_{e^2}$ may be the amplitude among 4 charged Goldstone bosons, which do not have the shift symmetry breaking $\phi^4$ interaction; $\mc{A}_\lambda$ is the amplitude among 4 neutral scalars. 

We therefore formulate the {\bf principle of amplitude decomposition}: any amplitude $\mc{M}$ depending on an independent set of parameters $\{g_i|i=1,\dots,n\}$ can always be decomposed into a combination of amplitude components $\mc{A}_{i_1,\dots,i_k}$. Each component corresponds to a physical theory where only a specific subset of parameters $\{g_{i_1},\dots,g_{i_k}\}$ are non-zero. Due to the independence of the parameters, these amplitude components are all physically valid and respect all the symmetries of the theory, including the gauge invariance. 
Within the on-shell bootstrap program, if a complete set of these components is on-shell constructible, then the full amplitude with arbitrary parameters can be recovered as their linear combination—even if the full amplitude itself fails the standard constructibility tests. This principle dramatically expands the range of theories accessible to on-shell methods. 
In particular, it is extremely helpful to examine the decomposition of the amplitude in terms of the charge parameters it depends on. Below we demonstrate how this decomposition works in both abelian and non-abelian theories.

\subsubsection{Single Photon}

We first consider the charged Goldstone bosons interacting with one external $U(1)$ gauge boson, the photon, through the minimal coupling.
The strength of the minimal coupling depends on the charges $q$ of the scalars as
\eq{
    \mc{M}(\phi^{+q},\bar\phi^{-q};\gamma) = qe\epsilon^\mu(p_1-p_2)_\mu \ .
}
When a photon is attached to a general $n$-scalar amplitude via the minimal coupling, we would have an amplitude of the generic form
\eq{ 
    \mc{M}_{n+1}(\phi^{q_1},\phi^{q_2},\dots,\phi^{q_n},\gamma) = \sum_{i=1}^n q_i m_i \ . 
}
where the $m_i$ are the contributions proportional to the charges of the external particles.
In terms of Feynman diagrams, there could be diagram where the photon is attached to internal propagators or vertices with charge $q_I$, but due to the charge conservation of the subamplitude $q_I = \sum_{i\in I} q_i$, it can also be split and absorbed into $m_{i\in I}$. 
It is crucial to realize that, in the above, the charges $q_i$ as the coefficients of $m_i$ are not independent due to the constraint of total charge conservation $\sum_i q_i=0$, and equivalently each of $m_i$ is usually not gauge invariant. 
Therefore, the number of independent charge parameters for $\mc{M}_{n+1}$ is actually $n-1$.

To isolate the constructible pieces, we re-parameterize the amplitude by choosing a basis for the $(n-1)$-dimensional charge space $\{Q_i|i=1,\dots,n-1\}$, which consists of independent combinations of $\{q_i\}$ respecting the constraint $\sum_i q_i=0$. 
A convenient choice is $\{Q_i=q_1+q_2+\cdots+q_i|i\neq n\}$ which leads to the decomposition:
\eq{\label{eq:sub_amp_ex1q}
    \mc{M}_{n+1} = \sum_{i=1}^n q_i m_i = \sum_{i=1}^{n-1} Q_i (m_{i}-m_{i+1}) \equiv \sum_{i=1}^{n-1} Q_i \mc{A}_i \ .
}
This yields the gauge invariant component $\mc{A}_i \equiv m_i - m_n$ in this case, so that the decomposition works as $\mc{M}_{n+1} =\sum_i Q_i \mc{A}_i$.\footnote{We are using the same notation $\mc{A}$ as the partial amplitudes for the flavor multiplets, because they share the same spirit: the $\mc{A}$'s are not full amplitudes, but components of the amplitudes that we can compute individually and simplify the calculation.} 
Crucially, each amplitude component $\mc{A}_i$ has a precise physical interpretation: it represents a physical, gauge-invariant amplitude in an auxiliary theory where \textit{only} particles $i$ and $i+1$ carry opposite unit charges, and all others are neutral. Formally,
\eq{\label{eq:sub_amp_ex1}
\mc{A}_i = m_i - m_{i+1} \equiv \mc{M}(\phi_1^0,\dots,\phi_i^{+1},\phi_{i+1}^{-1},\dots,\phi_n^0;\gamma)\ .
}
This decomposition admits an intuitive circuit theory analogy: the flow of charge through the amplitude can be likened to current flowing through a junction. 
The total current can be decomposed into $n-1$ independent currents flowing from particle $i$ to particle $i+1$. The amplitude decomposes analogously:\\%[1pt]
\newcommand\ncoord[2][0,0]{%
    \tikz[remember picture,overlay]{\path (#1) coordinate (#2);}%
}
% Define a handful of special cases of \ncoord
\newcommand\ccoord[1]{\ncoord[0.5em,0.9em]{#1}}%
\newcommand\bcoord[1]{\ncoord[0.4em,-0.3em]{#1}}%
\eq{
    \mc{M}(\phi_1^{q_1},\dots,\phi_n^{q_n};\gamma) & = \mc{M}({\ccoord{a}}{\phi_1}, {\ccoord{b}}{\phi_2},\dots) + \mc{M}(\phi_1, {\ccoord{c}}{\phi_2}, {\ccoord{d}}{\phi_3} \dots) + \dots \\
    &= Q_1\mc{A}_1 + Q_2\mc{A}_2 + \dots
}
\tikz[overlay,remember picture] {
\draw[red,->] (a) -- ++(0,0.4em) -| (b)
        node[above, near start] {\footnotesize $Q_1$};
\draw[red,->] (c) -- ++(0,0.4em) -| (d)
        node[above, near start] {\footnotesize $Q_2$};} 
\!\!\!Therefore we achieve a critical goal: within each component $\mc{A}_i$, the Adler's zero condition is restored for all ``neutral'' scalars
\eq{
    \lim_{p_j\to 0} \mc{A}_i =0 \qquad \text{if $j\neq i,i+1$ , and $\phi_j$ is a Goldstone boson.}
}
Consequently, if we can construct each of the component $\mc{A}_i$ with recursion relations, we will be able to construct the full $\mc{M}$ with arbitrary parameters $\{Q_i\}$.

As an illustration, let's consider 5-point amplitudes containing one photon and four charged scalars  with charges $q_i$, $i=1,2,3,4$. Introducing a scalar quartic coupling as before, the amplitude can be computed as the sum of four Feynman diagrams
\[
\resizebox{1.5cm}{!}{
	\begin{tikzpicture}[baseline=(o.base)]
	\begin{feynman}[small]
		\vertex [blob] (o) {};
		\vertex [below left=of o] (a);
		\vertex [above left=of o] (b);
		\vertex [below right=of o] (c);
		\vertex [above right=of o] (d);
        \vertex [right=of o] (p);
		\diagram*{ (o) --[scalar] {(a),(b),(c),(d)}, (o) --[photon] (p) };
	\end{feynman}
	\end{tikzpicture}
}=
q_1\resizebox{1cm}{!}{
	\begin{tikzpicture}[baseline=(o.base)]
	\begin{feynman}[small]
		\vertex (o);
		\vertex [below left=of o] (a);
		\vertex [above left=of o] (b);
		\vertex [below right=of o] (c);
		\vertex [above right=of o] (d);
		\coordinate (v) at ($(o)!.5!(a)$);
		\vertex [below right= of v] (p);
		\diagram*{ (o) --[scalar] {(a),(b),(c),(d)}, (v) --[photon] (p) };
	\end{feynman}
	\end{tikzpicture}
}+
q_2\resizebox{1cm}{!}{
	\begin{tikzpicture}[baseline=(o.base)]
	\begin{feynman}[small]
		\vertex (o);
		\vertex [below left=of o] (a);
		\vertex [above left=of o] (b);
		\vertex [below right=of o] (c);
		\vertex [above right=of o] (d);
		\coordinate (v) at ($(o)!.5!(b)$);
		\vertex [above right= of v] (p);
		\diagram*{ (o) --[scalar] {(a),(b),(c),(d)}, (v) --[photon] (p) };
	\end{feynman}
	\end{tikzpicture}
}+
q_3\resizebox{1.2cm}{!}{
\begin{tikzpicture}[baseline=(o.base)]
	\begin{feynman}[small]
		\vertex (o);
		\vertex [below left=of o] (a);
		\vertex [above left=of o] (b);
		\vertex [below right=of o] (c);
		\vertex [above right=of o] (d);
		\coordinate (v) at ($(o)!.5!(d)$);
		\vertex [below right= of v] (p);
		\diagram*{ (o) --[scalar] {(a),(b),(c),(d)}, (v) --[photon] (p) };
	\end{feynman}
	\end{tikzpicture}
}+
q_4\resizebox{1.2cm}{!}{
\begin{tikzpicture}[baseline=(o.base)]
	\begin{feynman}[small]
		\vertex (o);
		\vertex [below left=of o] (a);
		\vertex [above left=of o] (b);
		\vertex [below right=of o] (c);
		\vertex [above right=of o] (d);
		\coordinate (v) at ($(o)!.5!(c)$);
		\vertex [above right= of v] (p);
		\diagram*{ (o) --[scalar] {(a),(b),(c),(d)}, (v) --[photon] (p) };
	\end{feynman}
\end{tikzpicture}
} ,
\]
\eq{
    \mc{M}(\{\phi_i^{q_i}(p_i)\} ;\gamma(k)) = e\lambda\sum_{i=1}^4 q_i\frac{\varepsilon\cdot p_i}{k\cdot p_i} \ .
}
Each term in the sum is not gauge invariant, while the sum is. Now we write the amplitude as the sum of three terms:
\eq{
    \mc{M}(\{\phi_i^{q_i}(p_i)\}
    ;\gamma(k)) &= e\lambda\sum_{i=1}^3\left(\sum_{j=1}^i q_j\right)\left(\frac{\varepsilon\cdot p_i}{k\cdot p_i} - \frac{\varepsilon\cdot p_{i+1}}{k\cdot p_{i+1}}\right) \equiv e\lambda\sum_{i=1}^3 Q_i \mc{A}_i \ . 
}
This is exactly the decomposition shown in eq.~\eqref{eq:sub_amp_ex1q} and eq.~\eqref{eq:sub_amp_ex1}, where $Q_i = \sum_{j=1}^i q_j$ and
\eqs{
    \mc{A}_1 &= \mc{M}(\phi_1^{+1},\phi_2^{-1},\phi_3^0,\phi_4^0;\gamma) = \frac{\varepsilon\cdot p_1}{k\cdot p_1} - \frac{\varepsilon\cdot p_{2}}{k\cdot p_{2}} = \frac{p_{2,\mu}p_{1,\nu}}{(k\cdot p_1)(k\cdot p_2)}(k^\mu\varepsilon^\nu - k^\nu\varepsilon^\mu) \ , \\
    \mc{A}_2 &= \mc{M}(\phi_1^0,\phi_2^{+1},\phi_3^{-1},\phi_4^0;\gamma) = \frac{\varepsilon\cdot p_2}{k\cdot p_2} - \frac{\varepsilon\cdot p_{3}}{k\cdot p_{3}} = \frac{p_{3,\mu}p_{2,\nu}}{(k\cdot p_2)(k\cdot p_3)}(k^\mu\varepsilon^\nu - k^\nu\varepsilon^\mu) \ , \\
    \mc{A}_3 &= \mc{M}(\phi_1^0,\phi_2^0,\phi_3^{+1},\phi_4^{-1};\gamma) = \frac{\varepsilon\cdot p_3}{k\cdot p_3} - \frac{\varepsilon\cdot p_{4}}{k\cdot p_{4}} = \frac{p_{4,\mu}p_{3,\nu}}{(k\cdot p_3)(k\cdot p_4)}(k^\mu\varepsilon^\nu - k^\nu\varepsilon^\mu) \ .
}
These are obviously gauge invariant amplitudes where the photon interacts only with a unit-charge current flowing between two of the particles.

\subsubsection{Multiple Photons}

The decomposition principle extends naturally to amplitudes with multiple photons. For two photons attached to an $n$-point amplitude, the result is quadratic in the charges. Using the same basis $\{Q_i|i=1,\dots,n-1\}$. for the independent charge parameters, a general decomposition exists:
\eq{\label{eq:2gammadecom}
    \mc{M}(\phi_1^{q_1},\dots,\phi_n^{q_n};\gamma_1,\gamma_2) = \sum_{i\le j}^{n-1}Q_iQ_j\widetilde{\mc{A}_{i,j}}\ .
}
where each component $\widetilde{\mc{A}_{i,j}}$ is gauge invariant.
Note that since the $Q_iQ_j$ term is equivalent to the $Q_jQ_i$ term, we demand that $i\le j$ in $\widetilde{\mc{A}_{i,j}}$ to avoid double counting.
However, the components $\widetilde{\mc{A}_{i,j}}$ are not themselves individual physical amplitudes.  In the single photon case, each $\mc{A}_i$ is constructible because it is a valid amplitude of some particular theory with $Q_i = 1$ while $Q_{j\neq i}=0$. Now that two charges are involved in each term, we can perform a change of basis to such a more physical set $\mc{A}_{i,j}$, defined as the amplitudes where specific charge flows $Q_{i,j}$ are turned on with unit strength:
\eq{
    \mc{A}_{i,j} &= \mc{M}|_{Q_{i,j}=1,Q_{k\neq i,j}=0} = \widetilde{\mc{A}_{i,i}} + \widetilde{\mc{A}_{i,j}} + \widetilde{\mc{A}_{j,j}} \quad \text{if}\ i\neq j\ ,\\
    \mc{A}_{i,i} &= \mc{M}|_{Q_i=1,Q_{k\neq i}=0} = \widetilde{\mc{A}_{i,i}} \ .
}
We refer to $\{\mc{A}_{i,j}\}$ as the {\bf charge basis}, as each element corresponds to a physical amplitude with a well-defined charge configuration and can be constructed independently.
In terms of the charge basis, the decomposition is
\eq{
    \mc{M}(\phi_1^{q_1},\dots,\phi_n^{q_n};\gamma_1,\gamma_2) &= \sum_i^{n-1} Q_i^2\widetilde{\mc{A}_{i,i}} + \sum_{i < j}^{n-1} Q_iQ_j \widetilde{\mc{A}_{i,j}} \\
    &= \sum_i^{n-1} Q_i^2\mc{A}_{i,i} + \sum_{i < j}^{n-1}Q_iQ_j \left(\mc{A}_{i,j} - \mc{A}_{i,i} - \mc{A}_{j,j}\right) \\
    &= \sum_i^{n-1} Q_i(2Q_i - Q_{\rm tot}) \mc{A}_{i,i} + \sum_{i < j}^{n-1} Q_iQ_j \mc{A}_{i,j} \ .
}
where $Q_{\rm tot} \equiv \sum_{i}^{n-1}Q_i$.
As an example, take $n=4$ and $\{q_i\} = \{2,-1,-1,0\}$, so that $\{Q_i\} = \{2,1,0\}$ and $Q_{\rm tot} = 3$. The amplitude can be decomposed as
\eq{
    \mc{M}(\phi_1^{+2},\phi_2^{-1},\phi_3^{-1},\phi^0;\gamma_1,\gamma_2) &= 2\mc{A}_{1,1} - \mc{A}_{2,2} + 2\mc{A}_{1,2} \ .
}

Suppose we adopt the definition in eq.~\eqref{eq:sub_amp_ex1q}, they can be defined as the following physical amplitudes
\eq{
    \mc{A}_{i,j} &= \mc{M}(\dots,\phi_i^{+1},\phi_{i+1}^{-1},\dots,\phi_j^{+1},\phi_{j+1}^{-1},\dots ; \gamma_1,\gamma_2) \quad \text{if}\ |i-j|>1\ ,\\ 
    \mc{A}_{i,i+1} &= \mc{M}(\dots,\phi_i^{+1},\phi_{i+1}^{0},\phi_{i+2}^{-1},\dots ; \gamma_1,\gamma_2) \ ,\\ 
    \mc{A}_{i,i} &= \mc{M}(\dots,\phi_i^{+1},\phi_{i+1}^{-1},\dots ; \gamma_1,\gamma_2) \ .
}
For the latter two, each component only involves two charged scalars, thus we still have $n-2$ scalars with potential soft behavior. Nevertheless, in the first case we have to compute an amplitude with 4 charged scalars, where we have fewer soft conditions to make use of. In general, to compute an amplitude with $m$ photons (minimally coupled), the necessary number of charged scalars for a generic amplitude component $\mc{A}_{i_1,i_2,\dots,i_m}$ would be $2m$.

\subsubsection{Non-Abelian Gauge Group}

The principle of amplitude decomposition generalizes elegantly to non-Abelian gauge groups such as $SU(N)$. In this context, the role of the Abelian charges $Q_i$ is played by the color flow, represented by contractions of fundamental and anti-fundamental color indices. 
We consider scalars in an arbitrary representation, which can be built from a rank-$(k,l)$ tensor $\phi^{a_1,\dots,a_k}_{b_1,\dots,b_l}$, where the upper (lower) indices transform under the (anti-)fundamental representation. For example, the adjoint representation is in the decomposition of $(1,1)$ tensor
\eq{
    \phi^a_b = \phi_{\rm singlet}\ \delta^a_b + \phi_{\rm adj}^A\ (T^A)^a_b \ .
}
To form an $SU(N)$ invariant amplitude, it is necessary that the fundamental and anti-fundamental indices contract in pairs, which is very similar to the charge currents we dealt with in the Abelian case. 
For instance, one particular ordering in the single trace of adjoint scalars can be decomposed as
\begin{figure}[tbp]
    \centering
\tikzset{every picture/.style={line width=0.75pt}} %set default line width to 0.75pt        
\begin{tikzpicture}[x=0.75pt,y=0.75pt,yscale=-1,xscale=1]
%uncomment if require: \path (0,306); %set diagram left start at 0, and has height of 306

%Straight Lines [id:da15743986111437414] 
\draw    (67.17,71.9) -- (93.67,71.02) ;
\draw [shift={(73.92,71.67)}, rotate = 358.12] [fill={rgb, 255:red, 0; green, 0; blue, 0 }  ][line width=0.08]  [draw opacity=0] (8.93,-4.29) -- (0,0) -- (8.93,4.29) -- cycle    ;
%Straight Lines [id:da7975651274817592] 
\draw    (40.67,72.77) -- (67.17,71.9) ;
\draw [shift={(47.42,72.54)}, rotate = 358.12] [fill={rgb, 255:red, 0; green, 0; blue, 0 }  ][line width=0.08]  [draw opacity=0] (8.93,-4.29) -- (0,0) -- (8.93,4.29) -- cycle    ;
%Shape: Spring [id:dp6648224369884538] 
\draw   (67.05,138.82) .. controls (68.58,139.15) and (70.12,140.27) .. (70.16,142.56) .. controls (70.24,147.14) and (64.15,147.24) .. (64.12,145.52) .. controls (64.09,143.81) and (70.18,143.7) .. (70.26,148.28) .. controls (70.34,152.86) and (64.25,152.96) .. (64.22,151.25) .. controls (64.19,149.53) and (70.28,149.42) .. (70.36,154) .. controls (70.44,158.58) and (64.35,158.68) .. (64.32,156.97) .. controls (64.29,155.25) and (70.38,155.15) .. (70.46,159.72) .. controls (70.54,164.3) and (64.44,164.41) .. (64.41,162.69) .. controls (64.38,160.97) and (70.48,160.87) .. (70.56,165.45) .. controls (70.64,170.02) and (64.54,170.13) .. (64.51,168.41) .. controls (64.48,166.69) and (70.58,166.59) .. (70.66,171.17) .. controls (70.67,171.97) and (70.5,172.63) .. (70.2,173.18) ;
%Straight Lines [id:da660715719997394] 
\draw    (68.5,173.56) -- (95,172.69) ;
\draw [shift={(75.25,173.34)}, rotate = 358.12] [fill={rgb, 255:red, 0; green, 0; blue, 0 }  ][line width=0.08]  [draw opacity=0] (8.93,-4.29) -- (0,0) -- (8.93,4.29) -- cycle    ;
%Straight Lines [id:da9282323100578321] 
\draw    (42,174.43) -- (68.5,173.56) ;
\draw [shift={(48.75,174.21)}, rotate = 358.12] [fill={rgb, 255:red, 0; green, 0; blue, 0 }  ][line width=0.08]  [draw opacity=0] (8.93,-4.29) -- (0,0) -- (8.93,4.29) -- cycle    ;
%Straight Lines [id:da26691456165987737] 
\draw  [dash pattern={on 4.5pt off 4.5pt}]  (67.17,71.9) -- (66.83,37.33) ;
%Straight Lines [id:da37382955943104945] 
\draw    (98.67,252.1) -- (44.67,252.43) ;
\draw [shift={(66.67,252.3)}, rotate = 359.65] [fill={rgb, 255:red, 0; green, 0; blue, 0 }  ][line width=0.08]  [draw opacity=0] (8.93,-4.29) -- (0,0) -- (8.93,4.29) -- cycle    ;

% Text Node
\draw (60,20.4) node [anchor=north west][inner sep=0.75pt]    {$A$};
% Text Node
\draw (61.33,122.07) node [anchor=north west][inner sep=0.75pt]    {$B$};
% Text Node
\draw (31,76.4) node [anchor=north west][inner sep=0.75pt]    {$a$};
% Text Node
\draw (95.67,74.42) node [anchor=north west][inner sep=0.75pt]    {$b$};
% Text Node
\draw (31.33,177.07) node [anchor=north west][inner sep=0.75pt]    {$a$};
% Text Node
\draw (97,176.09) node [anchor=north west][inner sep=0.75pt]    {$b$};
% Text Node
\draw (111,38.4) node [anchor=north west][inner sep=0.75pt]  [font=\small]  {$=\left( T^{A}\right)_{b}^{a}$};
% Text Node
\draw (110.67,140.4) node [anchor=north west][inner sep=0.75pt]  [font=\small]  {$=\left( T^{B}\right)_{b}^{a}$};
% Text Node
\draw (180.33,142) node [anchor=north west][inner sep=0.75pt]  [font=\small] [align=left] {for gauge boson ;};
% Text Node
\draw (179.67,40.5) node [anchor=north west][inner sep=0.75pt]  [font=\small] [align=left] {for Goldstone boson ; };
% Text Node
\draw (32,253.73) node [anchor=north west][inner sep=0.75pt]    {$b$};
% Text Node
\draw (100.67,255.5) node [anchor=north west][inner sep=0.75pt]    {$a$};
% Text Node
\draw (127,239.73) node [anchor=north west][inner sep=0.75pt]    {$=\delta _{a}^{b}$};

\end{tikzpicture}

    \caption{Here, we use diagrams to represent the group tensors. To make the distinction clearer, we use dashed lines to represent the adjoint representation index $A$ of the Goldstone boson and looped lines to denote the adjoint representation index $B$ of the gauge boson. }
    \label{fig:tensors}
\end{figure}
\eqs{
    &\mbox{tr}(T^{A_1}T^{A_2}\dots T^{A_n}) = (T^{A_1})^{a_1}_{b_1}(T^{A_2})^{a_2}_{b_2}\dots (T^{A_n})^{a_n}_{b_n} \times \delta^{b_1}_{a_2}\delta^{b_2}_{a_3}\dots\delta^{b_n}_{a_1}\ ,\label{eq:Adelta}\\
    &\begin{tikzpicture}[x=0.75pt,y=0.75pt,yscale=-1,xscale=1]
%uncomment if require: \path (0,171); %set diagram left start at 0, and has height of 171

%Shape: Arc [id:dp3391283583442479] 
\draw  [draw opacity=0][dash pattern={on 4.5pt off 4.5pt}] (95.13,95.51) .. controls (90.37,95.71) and (85.62,93.97) .. (82.17,90.31) .. controls (79.2,87.16) and (77.72,83.15) .. (77.67,79.07) -- (95.02,78.21) -- cycle ; \draw [dash pattern={on 4.5pt off 4.5pt}] [dash pattern={on 4.5pt off 4.5pt}]  (95.13,95.51) .. controls (90.37,95.71) and (85.62,93.97) .. (82.17,90.31) .. controls (79.2,87.16) and (77.72,83.15) .. (77.67,79.07) ;  
%Shape: Arc [id:dp9074773432913286] 
\draw  [draw opacity=0] (112.02,74.09) .. controls (113.21,79.83) and (111.39,86.16) .. (106.68,90.59) .. controls (104.13,93) and (101.06,94.53) .. (97.89,95.17) -- (95.02,78.21) -- cycle ; \draw    (112.37,77.14) .. controls (112.49,81.99) and (110.57,86.93) .. (106.68,90.59) .. controls (104.13,93) and (101.06,94.53) .. (97.89,95.17) ;  \draw [shift={(112.02,74.09)}, rotate = 98.47] [fill={rgb, 255:red, 0; green, 0; blue, 0 }  ][line width=0.08]  [draw opacity=0] (8.93,-4.29) -- (0,0) -- (8.93,4.29) -- cycle    ;
%Shape: Arc [id:dp6397147186570572] 
\draw  [draw opacity=0] (88.63,62.35) .. controls (95.16,59.53) and (102.88,60.8) .. (107.87,66.1) .. controls (111.94,70.42) and (113.23,76.38) .. (111.84,81.87) -- (95.02,78.21) -- cycle ; \draw    (91.53,61.38) .. controls (97.34,59.99) and (103.6,61.57) .. (107.87,66.1) .. controls (111.94,70.42) and (113.23,76.38) .. (111.84,81.87) ;  \draw [shift={(88.63,62.35)}, rotate = 356.26] [fill={rgb, 255:red, 0; green, 0; blue, 0 }  ][line width=0.08]  [draw opacity=0] (8.93,-4.29) -- (0,0) -- (8.93,4.29) -- cycle    ;
%Shape: Arc [id:dp4329940479438681] 
\draw  [draw opacity=0] (78.11,82.76) .. controls (76.74,76.9) and (78.53,70.37) .. (83.36,65.82) .. controls (88.42,61.05) and (95.46,59.75) .. (101.42,61.88) -- (95.02,78.21) -- cycle ; \draw    (77.69,79.73) .. controls (77.43,74.74) and (79.35,69.6) .. (83.36,65.82) .. controls (88.42,61.05) and (95.46,59.75) .. (101.42,61.88) ;  \draw [shift={(78.11,82.76)}, rotate = 276.93] [fill={rgb, 255:red, 0; green, 0; blue, 0 }  ][line width=0.08]  [draw opacity=0] (8.93,-4.29) -- (0,0) -- (8.93,4.29) -- cycle    ;
%Shape: Ellipse [id:dp7792902171947688] 
\draw  [fill={rgb, 255:red, 0; green, 0; blue, 0 }  ,fill opacity=1 ] (82.01,112.54) .. controls (81.21,111.69) and (81.26,110.35) .. (82.13,109.54) .. controls (82.99,108.72) and (84.33,108.75) .. (85.13,109.6) .. controls (85.93,110.44) and (85.87,111.78) .. (85.01,112.6) .. controls (84.15,113.41) and (82.8,113.38) .. (82.01,112.54) -- cycle ;
%Shape: Ellipse [id:dp4312261279793421] 
\draw  [fill={rgb, 255:red, 0; green, 0; blue, 0 }  ,fill opacity=1 ] (102.37,112.94) .. controls (101.57,112.1) and (101.62,110.75) .. (102.49,109.94) .. controls (103.35,109.13) and (104.69,109.15) .. (105.49,110) .. controls (106.28,110.85) and (106.23,112.19) .. (105.37,113) .. controls (104.51,113.81) and (103.16,113.79) .. (102.37,112.94) -- cycle ;
%Shape: Ellipse [id:dp3704319933535024] 
\draw  [fill={rgb, 255:red, 0; green, 0; blue, 0 }  ,fill opacity=1 ] (92.63,114.28) .. controls (91.84,113.43) and (91.89,112.09) .. (92.75,111.27) .. controls (93.62,110.46) and (94.96,110.49) .. (95.76,111.33) .. controls (96.55,112.18) and (96.5,113.52) .. (95.64,114.34) .. controls (94.77,115.15) and (93.43,115.12) .. (92.63,114.28) -- cycle ;
%Straight Lines [id:da9891483957737989] 
\draw    (256,68.43) -- (255.68,102.38) ;
\draw [shift={(255.79,90.41)}, rotate = 270.55] [fill={rgb, 255:red, 0; green, 0; blue, 0 }  ][line width=0.08]  [draw opacity=0] (8.93,-4.29) -- (0,0) -- (8.93,4.29) -- cycle    ;
%Straight Lines [id:da8428654588966698] 
\draw    (303.33,59.77) -- (264.73,60.16) ;
\draw [shift={(279.03,60.02)}, rotate = 359.41] [fill={rgb, 255:red, 0; green, 0; blue, 0 }  ][line width=0.08]  [draw opacity=0] (8.93,-4.29) -- (0,0) -- (8.93,4.29) -- cycle    ;
%Straight Lines [id:da3412952386758935] 
\draw  [dash pattern={on 4.5pt off 4.5pt}]  (82.8,66.71) -- (62.67,45.1) ;
%Straight Lines [id:da6869610835514258] 
\draw  [dash pattern={on 4.5pt off 4.5pt}]  (125.88,110.33) -- (106.37,89.97) ;
%Straight Lines [id:da0015598915739611918] 
\draw  [dash pattern={on 4.5pt off 4.5pt}]  (108.25,67.22) -- (128.98,47.14) ;
%Straight Lines [id:da06242530425944404] 
\draw  [dash pattern={on 4.5pt off 4.5pt}]  (60.6,112.09) -- (82.92,90.52) ;
%Straight Lines [id:da88184111557188] 
\draw    (309.33,68.43) -- (309.01,102.38) ;
\draw [shift={(309.23,78.91)}, rotate = 90.55] [fill={rgb, 255:red, 0; green, 0; blue, 0 }  ][line width=0.08]  [draw opacity=0] (8.93,-4.29) -- (0,0) -- (8.93,4.29) -- cycle    ;
%Shape: Ellipse [id:dp5313319947567048] 
\draw  [fill={rgb, 255:red, 0; green, 0; blue, 0 }  ,fill opacity=1 ] (264.86,108.28) .. controls (264.32,106.91) and (265.02,105.38) .. (266.43,104.85) .. controls (267.85,104.32) and (269.43,105.01) .. (269.98,106.37) .. controls (270.52,107.74) and (269.82,109.28) .. (268.4,109.8) .. controls (266.99,110.33) and (265.4,109.65) .. (264.86,108.28) -- cycle ;
%Shape: Ellipse [id:dp34631741696357843] 
\draw  [fill={rgb, 255:red, 0; green, 0; blue, 0 }  ,fill opacity=1 ] (278.19,108.95) .. controls (277.65,107.58) and (278.35,106.04) .. (279.77,105.52) .. controls (281.18,104.99) and (282.77,105.67) .. (283.31,107.04) .. controls (283.85,108.41) and (283.15,109.94) .. (281.74,110.47) .. controls (280.32,111) and (278.74,110.31) .. (278.19,108.95) -- cycle ;
%Shape: Ellipse [id:dp12057450617883958] 
\draw  [fill={rgb, 255:red, 0; green, 0; blue, 0 }  ,fill opacity=1 ] (292.86,108.28) .. controls (292.32,106.91) and (293.02,105.38) .. (294.43,104.85) .. controls (295.85,104.32) and (297.43,105.01) .. (297.98,106.37) .. controls (298.52,107.74) and (297.82,109.28) .. (296.4,109.8) .. controls (294.99,110.33) and (293.4,109.65) .. (292.86,108.28) -- cycle ;

% Text Node
\draw (39.33,30.34) node [anchor=north west][inner sep=0.75pt]  [color={rgb, 255:red, 0; green, 0; blue, 0 }  ,opacity=1 ]  {$A_{1}$};
% Text Node
\draw (131.06,30.31) node [anchor=north west][inner sep=0.75pt]    {$A_{2}$};
% Text Node
\draw (128.52,110.09) node [anchor=north west][inner sep=0.75pt]    {$A_{3}$};
% Text Node
\draw (38.34,109.8) node [anchor=north west][inner sep=0.75pt]    {$A_{n}$};
% Text Node
\draw (241.69,43.41) node [anchor=north west][inner sep=0.75pt]  [color={rgb, 255:red, 74; green, 144; blue, 226 }  ,opacity=1 ]  {$1$};
% Text Node
\draw (308.71,43) node [anchor=north west][inner sep=0.75pt]  [color={rgb, 255:red, 74; green, 144; blue, 226 }  ,opacity=1 ]  {$2$};
% Text Node
\draw (260.43,48.14) node [anchor=north west][inner sep=0.75pt]  [font=\scriptsize]  {$b_{1}$};
% Text Node
\draw (241.25,63.27) node [anchor=north west][inner sep=0.75pt]  [font=\scriptsize]  {$a_{1}$};
% Text Node
\draw (242.94,109.99) node [anchor=north west][inner sep=0.75pt]  [color={rgb, 255:red, 74; green, 144; blue, 226 }  ,opacity=1 ]  {$n$};
% Text Node
\draw (240.77,95.5) node [anchor=north west][inner sep=0.75pt]  [font=\scriptsize]  {$b_{n}$};
% Text Node
\draw (310.72,95.91) node [anchor=north west][inner sep=0.75pt]  [font=\scriptsize]  {$a_{3}$};
% Text Node
\draw (293.83,49.5) node [anchor=north west][inner sep=0.75pt]  [font=\scriptsize]  {$a_{2}$};
% Text Node
\draw (310.14,62.21) node [anchor=north west][inner sep=0.75pt]  [font=\scriptsize]  {$b_{2}$};
% Text Node
\draw (308.61,107.65) node [anchor=north west][inner sep=0.75pt]  [color={rgb, 255:red, 74; green, 144; blue, 226 }  ,opacity=1 ]  {$3$};
% Text Node
\draw (184.8,74.8) node [anchor=north west][inner sep=0.75pt]    {$\sim $};

\end{tikzpicture} \label{eq:Adelta1}
}
where each of the $\delta_{a_i}^{b_j}$ represents a contraction from the $i$th to the $j$th particle, similar to a ``charge current'' flowing between them\footnote{
The Levi-Civita $\epsilon$ tensors, which can also form $SU(N)$ singlets, can be interpreted by converting a flow into its dual form.
For instance, an $SU(3)$ tensor $\epsilon^{a_1 a_2 a_3}$ turns the out-flow of particle 3 into two in-flows as $\delta^{a_1}_{b_1}\delta^{a_2}_{b_2} - \delta^{a_1}_{b_2}\delta^{a_2}_{b_1}$. This provides a universal picture of color flows for any $SU(N)$ invariant amplitude.}.
This perspective leads to a powerful graphical representation with the rules in figure~\ref{fig:tensors} and an example in eq.~\eqref{eq:Adelta1}. 
A particular color structure can always be described by a specific pattern of index contractions, which we call a "current configuration" and denote by $\mc{C}$. The full amplitude is then a sum over all such valid configurations $\mc{C}$:
\eq{\label{eq:decom_part}
    \mc{M}(1^{A_1},2^{A_2},\dots,n^{A_n}) = (T^{A_1})^{a_1}_{b_1}(T^{A_2})^{a_2}_{b_2}\dots (T^{A_n})^{a_n}_{b_n} \times \sum_{\cal C} \mc{M}^{b_1\cdots b_n}_{a_1\cdots a_n}({\cal C})  %(1^{a_1}_{b_1},\dots,n^{a_n}_{b_n})
}
For instance, we have $\mc{M}^{b_1\cdots b_n}_{a_1\cdots a_n}({\cal C}) \sim \delta^{b_1}_{a_2}\delta^{b_2}_{a_3}\dots\delta^{b_n}_{a_1}$ for the configuartion $\mc{C}$ depicted in eq.~\eqref{eq:Adelta}. 

A key strength of this interpretation is its generality. The ``current configurations'' $\mc{C}$ are defined by the pattern of fundamental and anti-fundamental index contractions in the $SU(N)$-invariant tensor of the amplitude. This formulation is not limited to scalars in the adjoint representation, but applies universally to scalars in any representation that can be built from generic tensor representations. The ``current'' interpretation of the color structure reflects the $SU(N)$ invariance in the same way that the counterpart in the Abelian theory, which is equivalent to the charge conservation, reflects the $U(1)$ invariance, while both are actually the conservation laws dictated by the asymptotic global gauge symmetry.

Now we try to add a gluon $G$ to the amplitude for each of the configuration $\mc{C}$
\eq{
    \mc{M}^{b_1\cdots}_{a_1\cdots}(\mc{C}) \quad\longrightarrow\quad \mc{M}^{b_1\cdots}_{a_1\cdots}(\mc{C};G^B) \ .
}
Consider each current in $\mc{C}$ is charged under a different $SU(N)$ group, so that the symmetry is fictitiously extended to $SU(N)^{n_c}$, where $n_c$ is the number of currents uniquely determined by the representations of the scalars and is thus universal for all $\mc{C}$. For example, in the charge configuration eq.~\eqref{eq:Adelta}, there are exactly $n_c=n$ currents, while a current $\delta_{a_{i}}^{b_{i-1}}$ belongs to the group $SU(N)_i$ for which the $\phi_i$ is under the fundamental representation and $\phi_{i-1}$ is under the anti-fundamental representation and all the others are singlets. 
Afterwards, we may first gauge the $n$ groups by independent couplings $\{g_i\}$. This structure is directly reminiscent of a quiver gauge theory \cite{Douglas:1996sw}. In this auxiliary theory, the scalars are bi-fundamental fields linking a product of $SU(N)$ gauge groups. The component $\mc{A}_i$ is then the amplitude in a specific quiver where only the $i$-th gauge group is dynamically gauged. This perspective solidifies the physical interpretation of our decomposition: we are dissecting the original amplitude into contributions from simpler, auxiliary quiver gauge theories.. When the independent gauge couplings are identified $g_i=g$, we return to the original theory.

The amplitude with one gauge boson would be linear in $g_i$ and can be written as $\mc{M}^{b_1\cdots}_{a_1\cdots}(\mc{C};G^B) = \sum_i g_i\mc{M}_i{}^{b_1\cdots}_{a_1\cdots}(\mc{C};G^B)$. In each component $\mc{M}_i{}^{b_1\cdots}_{a_1\cdots}(\mc{C};G^B)$, the $SU(N)_i$ invariance requires the invariant tensor $(T^B)_{a_i}^{b_{i-1}}$, while the other $SU(N)_{j\neq i}$ symmetries require a factor of $\delta_{a_j}^{b_{j-1}}$ which remains the same as in $\mc{M}(\mc{C})$. Therefore, the amplitude of a particular current configuration $\mc{C}$ with a gauge boson can be decomposed as
\eqs{
    & \mc{M}^{b_1\cdots}_{a_1\cdots}(\mc{C};G^B) = \sum_{i=1}^{n_c} g_i \times \delta_{a_2}^{b_1}\cdots (T^B)_{a_i}^{b_{i-1}} \cdots \delta_{a_1}^{b_n} \times \mc{A}_i(\mc{C};G^B) \label{eq:decom_NA}\\
    & \input{Fig_decomposeNA}  \label{fig:decom_NA}
}
where $\mc{A}_i$ are the amplitude components that only depend on the kinematic variables. For a general current configuration, we simply replace the factor $\delta_{a_i}^{b_j}$ corresponding to the current charged under the gauged group by $(T^B)_{a_i}^{b_j}$, and keep all the other $\delta$'s for the ungauged group. 
In eq.~\eqref{fig:decom_NA}, we illustrate the process diagramatically, where the direct sum signifies a sum over the amplitude components with each of the flavor tensor instead of a sum over the flavor tensors themselves.
Since the gauge boson only interacts with two of the scalars in each component, the Adler's zero condition for all the other scalars is hence restored. Similar to the Abelian cases, these components are all physically valid amplitudes with gauge invariance, for a fictitious theory of $SU(N)^n$ symmetry where one of them is gauged, therefore we expect that $\mc{A}_i$ may be individually constructed by recursion relations. Finally, when the gauge couplings are identified $g_i=g$, we would get the original amplitude we need. 

A discerning reader may note that for the adjoint representation, the single-trace structures we use form an over-complete set due to trace identities. This redundancy, however, does not impede our recursive algorithm. The decomposition in eq.~\eqref{eq:decom_NA} is applied linearly to the invariant tensor of a given configuration. Our recursive construction does not require solving for the components $\mc{A}_i$ from a known full amplitude; rather, it builds them directly from lower-point amplitudes and soft theorems. The consistency of the recursion -- enforced by unitarity, gauge invariance, and the soft behavior of neutral scalars -- ensures that the resulting components for any valid configuration $\mc{C}$ are physically sound. The final physical amplitude, obtained by summing over a complete set of configurations, automatically projects onto the correct $SU(N)$ structure, with unphysical $U(1)$ components decoupling \cite{Elvang:2013cua}. The fundamental index contraction picture provides the most transparent and general interpretation of the 'charge flows' we are gauging individually.

To illustrate the whole process, we can apply the algorithm to the simplest Yang-Mills amplitude, the 4-point MHV amplitude $\mc{M}_4(1^-,2^-,3^+,4^+)$. Suppose the 4th gluon is added to the 3-point amplitude, which can be decomposed into ``current configurations'' as
\eq{
    \mc{M}_3(1^-,2^-,3^+) = g\left(\tr[123]-\tr[132]\right) \frac{\vev{12}^3}{\vev{13}\vev{23}}
}
where the traces $\tr[i_1,\dots,i_n]\equiv \tr(T^{A_{i_1}}\dots T^{A_{i_n}})$ are simply single-loop currents. The decomposition eq.~\eqref{eq:decom_NA} applies to the two terms independently:
\eq{
    &\mc{M}(\mc{C}_1;4^+) = g \left(\tr[1423] \mc{A}_{(12)}(\mc{C}_1;4^+) + \tr[1243] \mc{A}_{(23)}(\mc{C}_1;4^+) + \tr[1234] \mc{A}_{(31)}(\mc{C}_1;4^+) \right) \ ,\\
    &\mc{M}(\mc{C}_2;4^+) = -g \left(\tr[1432] \mc{A}_{(13)}(\mc{C}_2;4^+) + \tr[1342] \mc{A}_{(32)}(\mc{C}_2;4^+) + \tr[1324] \mc{A}_{(21)}(\mc{C}_2;4^+) \right) \ .
}
The amplitude components $\mc{A}_{(ij)}$ denote that the gluon $4^+$ acts on the current flowing from $i$ to $j$. In this case, it can be directly given by multiplying a soft factor $\frac{\vev{ij}}{\vev{4i}\vev{4j}}$ derived in eq.~\eqref{eq:soft_expansion}, where the gauge coupling $g$ and the generator $T^{A_4}$ are taken out of $\mc{A}_i$ as shown in eq.~\eqref{eq:decom_NA}.
For example, we have
\eqs{
    \mc{A}_{(12)}(\mc{C}_1;4^+) &= g\frac{\vev{12}}{\vev{14}\vev{24}}\frac{\vev{12}^3}{\vev{13}\vev{23}} = g\frac{\vev{12}^4}{\vev{14}\vev{42}\vev{23}\vev{31}} \ , \\
    \mc{A}_{(31)}(\mc{C}_1;4^+) &= g\frac{\vev{31}}{\vev{34}\vev{14}}\frac{\vev{12}^3}{\vev{13}\vev{23}} = g\frac{\vev{12}^4}{\vev{12}\vev{23}\vev{34}\vev{41}} \ ,\dots
}
All the six terms add up according to eq.~\eqref{eq:decom_part} to the result exactly matching with the Parke-Taylor formula
\eq{
    \mc{M}_4(1^-,2^-,3^+,4^+) = \sum_{\mc{C}}\mc{M}(\mc{C};4^+) = g^2 \tr[1234] \frac{\vev{12}^4}{\vev{12}\vev{23}\vev{34}\vev{41}} + \text{permutations}
}

In this way, we derive the non-Abelian version of the Amplitude Decomposition. We find that the common feature of both decompositions is to find the constituting currents of the interaction, and consider their individual interactions with the gauge boson. In the Abelian case, the strengths of the currents are characterized by the independent charge parameters $Q_i$, and the charge conservation implies that there are $n-1$ independent currents. In the non-Abelian cases, the currents are the $\delta$ contractions between the (anti-)fundamental indices of the particles. The rest of the decompositions in both cases are actually analogous. With the above in mind, it's important to acknowledge that the decomposition we present here apply primarily at tree-level, where the charges of the particles uniquely determine the charges of all internal lines. When loop-level contributions are considered, however, the decomposition would depend on the charges of the particles running in the loop. We will be studying the loop level decomposition and its implication in the future work.

\subsection{Internal Photons}\label{sec:int_ph}

A final challenge involves amplitudes with internal photons, whose contributions are genuinely ambiguous from a purely on-shell, unitarity-based perspective. For instance, the photon-exchange diagram in a four-scalar amplitude is determined only up to a contact term by factorization alone.
Even if we know that the contact term is absent for Goldstone bosons, we cannot use the Adler's zero condition to eliminate it, as the photon exchange amplitude already violates the Adler's zero condition -- it turns out that this contribution is irrespective of whether the scalars are Goldstones. Therefore, we have to come up with a new principle to fix the photon exchange contribution. 

To resolve this ambiguity, we introduce a new on-shell principle: the contribution from an internal photon is characterized by the total angular momentum $J=1$ in this channel. This allows us to identify the correct contribution via its unique contraction pattern in the spinor-helicity formalism, as detailed in \cite{Jiang:2020rwz,Shu:2021qlr}.
For example, the four-scalar amplitude with a photon exchanged between $\{\phi_1,\phi_2\}$ and $\{\phi_3,\phi_4\}$ can be written as
\eq{
    \mc{A}_{e^2}(1^{+q},2^{-q},3^{+q'},4^{-q'}) &= qq' \frac{e^2}{s_{12}} (s_{13}-s_{14}) = qq'e^2\frac{\vev{13}\vev{24}+\vev{14}\vev{23}}{\vev{12}\vev{34}} \\
    &= qe\frac{\ket{1}_\alpha\ket{2}_\beta}{\vev{12}} \times (\epsilon^{\alpha\alpha'}\epsilon^{\alpha\beta'}+\epsilon^{\alpha\beta'}\epsilon^{\beta\alpha'}) \times q'e\frac{\ket{3}_{\alpha'}\ket{4}_{\beta'}}{\vev{34}} \ ,
}
where the spinor contractions appear symmetric in the numerator. The reason for this particular pattern can be seen in the center of mass frame, where the amplitude has the angular distribution of $P_1(\cos\theta)=\cos\theta$ indicating $J=1$. The anti-symmetric contraction would instead provide a $J=0$ distribution
\eq{
    \frac{\ket{1}_\alpha\ket{2}_\beta}{\vev{12}} \times (\epsilon^{\alpha\alpha'}\epsilon^{\alpha\beta'} -\epsilon^{\alpha\beta'}\epsilon^{\beta\alpha'}) \times \frac{\ket{3}_{\alpha'}\ket{4}_{\beta'}}{\vev{34}} = 1 = P_0(\cos\theta)
}
which may come from a contact term $\lambda\phi^4$. The correspondence between the spinor contraction pattern and the angular momentum can be easily derived from the decomposition of tensor representation $\mathbf{2\times 2 = 1+3}$, while the rigorous proof can be found in \cite{Arkani-Hamed:2017jhn,Jiang:2020rwz}. 

This angular momentum selection rule serves a different purpose than the first principles of soft theorems or Adler's zero. It acts as a definitional scheme to resolve an inherent on-shell ambiguity. The decomposition in eq.~\eqref{eq:amp_scalar_QED}, $\mc{M}=e^2\mc{A}_{e^2}+\lambda\mc{A}_\lambda$, is not uniquely fixed by factorization alone, as the contact term $\mc{A}_\lambda$ does not contribute to any pole residues. The $J=1$ rule provides a specific prescription: it assigns to the photon exchange term $\mc{A}_{e^2}$ the component that transforms in the $J=1$ partial wave, effectively projecting out the $J=0$ contact interaction. This scheme robustly matches the result derived from Feynman diagrams, where the photon propagator inherently acts as a $J=1$ projector. While other subtraction schemes are conceivable, the  $J=1$ rule is physically well-motivated and ensures consistency with standard perturbative calculations.

To demonstrate how it works in the gauged NLSM, we consider $\mc{M}(1^{+q},2^{-q},3^{+q'}, 4^{-q'},\gamma)$, the internal photon contribution to the 5-point amplitude that is proportional to $e^3$. 
There are two contributions related to the internal photon with different poles, $s_{12}=0$ and $s_{34}=0$, which are both gauge independent. 
Without loss of generality, we compute the diagram where the internal photon is exchanged between the particles $\{\phi_1,\phi_2,\gamma\}$ and $\{\phi_3,\phi_4\}$, whose residue can be computed by unitarity
\eq{
    {\rm Res}_{s_{34}=0}\mc{A}_{e^3}(1^{+q},2^{-q},3^{+q'},4^{-q'},\gamma) = \sum_h \mc{A}_L(1^{+q},2^{-q},\gamma^{(h)},\gamma'^{(h')}) \times \mc{A}_R(3^{+q'},4^{-q'},\gamma'^{(-h')}) \ .
}
Suppose the external photon $\gamma$ carries helicity $h=+1$. In this case, only the $h=-1$ component gives a non-vanishing contribution to $\mathcal{M}_L$, owing to the helicity-selection rules \cite{Mangano:1990by}: same-helicity vector amplitudes vanish in massless renormalizable interactions. The non-zero amplitude is then
\begin{align}
\mathcal{A}_L(1^{+q},2^{-q},\gamma^{(+)},\gamma'^{(-)})
&= q^2 e^2 \frac{\langle 1\gamma'\rangle \langle 2\gamma'\rangle}{\langle 1\gamma\rangle \langle 2\gamma\rangle}.
\end{align}
In contrast, the right-handed component reads
\begin{align}
\mathcal{A}_R(3^{+q'},4^{-q'},\gamma'^{(+)})
&= q' e \frac{[3\gamma'][4\gamma']}{[34]}.
\end{align}
Note that the two subamplitudes are only defined for complex shifted momenta with on-shell $\gamma'$ at the pole $s_{34}=0$.
When multiplying them together, there are two different ways to combine the helicity spinors:
\eq{
    \lim_{s_{34}=0}\mc{A}_L\times\mc{A}_R 
    &= q^2q'e^3\frac{\bra{1}(p_3+p_4)|3]\bra{2}(p_3+p_4)|4]}{\vev{1\gamma}\vev{2\gamma}[34]} = q^2q'e^3\frac{[43]}{\vev{1\gamma}\vev{2\gamma}} \vev{14}\vev{23} \\
    &= q^2q'e^3\frac{\bra{1}(p_3+p_4)|4]\bra{2}(p_3+p_4)|3]}{\vev{1\gamma}\vev{2\gamma}[34]} = q^2q'e^3\frac{[43]}{\vev{1\gamma}\vev{2\gamma}} \vev{13}\vev{24}\ .
}
So the unitarity principle only fixes the amplitude to a combination of the two terms. However, observe that only the symmetric contraction leads to the $J=1$ partial wave:
\eq{
    \frac{1}{s_{34}}\times q^2q'e^3\frac{[43]}{\vev{1\gamma}\vev{2\gamma}} (\vev{14}\vev{23} + \vev{13}\vev{24}) \ ,
}
Adding the second contribution from the pole at $s_{12}=0$, the entire $e^3$ internal photon contribution is given by
\eq{
    \mc{A}_{e^3}(1^{+q},2^{-q},3^{+q'},4^{-q'},\gamma) = qq'e^3\left[q\frac{\vev{13}\vev{24}+\vev{14}\vev{23}}{\vev{1\gamma}\vev{2\gamma}\vev{34}} + q'\frac{\vev{13}\vev{24}+\vev{14}\vev{23}}{\vev{12}\vev{3\gamma}\vev{4\gamma}}\right] \ ,
}
which reproduces the Feynman diagram calculation.

\section{Gauged Soft Recursion}

We are now ready to extend the original soft recursion to include gauge interactions. As discussed earlier, the presence of gauge interactions breaks the shift symmetry acting on the charged Goldstones, preventing them from behaving as strictly soft. Actually, as is well known, these charged scalars acquire loop-induced masses $\Delta m^2\propto \alpha/4\pi$ and become pseudo-Nambu-Goldstone bosons. 
In this work, we focus on tree-level amplitudes and treat the scalars as massless. While our results can be promoted to the massive case using massive spinor variables \cite{Arkani-Hamed:2017jhn}, potential ambiguities in this promotion\footnote{
Such as $\lambda_\alpha\lambda_\beta\tilde\lambda_{\dot\alpha}\tilde\lambda_{\dot\beta} \to A=p_{\alpha\dot\alpha}p_{\beta\dot\beta}$ or $B=p_{\alpha\dot\beta}p_{\beta\dot\alpha}$, while they differ by $A-B=m^2\epsilon_{\alpha\beta}\epsilon_{\dot\alpha\dot\beta}$.
} can be resolved by enforcing the correct factorization properties on poles, a technical point we will not elaborate on here.

The goal is to construct the amplitude componnet $\mc{A}_{i_1,\dots,i_l}$ from the charge decomposition of $\mc{M}_{n+l}$ with $n$ Goldstone bosons and $l$ gauge bosons. The gauged soft recursion modifies the standard approach in two ways. First, the soft factor $F(z)$ now includes factors $(1-a_s z)$ for each of the $l$ gauge bosons, in addition to the factors from the $n_*$ neutral scalars, yielding 
\begin{equation}
    F(z)\sim z^{n_*\sigma+l}\quad ,\quad n_*=\max(n-2l,0) = \text{ number of soft scalars} \ .
\end{equation}
Second, the large-$z$ behavior of the amplitude is softened by the leading soft factor of each gauge boson $S^{(0)}\sim p_s^{-1} \sim z^{-1}$, so that
\eq{
    \lim_{z\to\infty} \hat{\mc{A}}_{i_1,\dots,i_l}(z) \sim z^{m-l} \ ,
}
where $m$ is the power of momenta from the soft blocks. The condition for the boundary term $B_{\infty}$ to vanish therefore becomes
\eq{\label{eq:Bcondition}
    \lim_{z\to\infty}\frac{\hat{\mc{A}}_{i_1,\dots,i_l}(z)}{F(z)} \sim z^{m-2l-n_*\sigma} \qquad \Rightarrow\quad m-2l-n_*\sigma < 0
}
Following \cite{Cheung:2015ota}, we characterize the scalar EFTs by the parameter $\rho = (m-2)/(n-2)$, which remains invariant under recursion. For an amplitude with a single photon ($l=1$), the constructibility condition becomes $\rho<\sigma$. 
To ensure constructibility with an arbitrary number of photons, we require the stronger condition $\rho<1$, which is satisfied by the Non-Linear Sigma Model (NLSM) with $(\rho,\sigma)=(0,1)$, the primary focus of this paper.

With the boundary term eliminated, we can apply Cauchy's theorem to construct the amplitude components. The function $\hat{\mc{A}}_{i_1,\dots,i_l}(z)/F(z)$ has poles from two sources: 
\begin{enumerate}
    \item The propagators of the hard factorization channels $z=z_I^\pm$ which appeared in the original soft recursion relation.

    \item The soft limits of the gauge bosons $z=1/a_s,\,s=1,\dots,l$. These arise from diagrams where a gauge boson is attached an external scalar leg\footnote{
    Under the all-line shift, the propagator $1/(\hat{p}_i\cdot \hat{p}_s)$ vanishes at $z=1/a_s$. The poles at $z=1/a_i$ from the scalar soft limits are canceled by corresponding zeros in the 3-point minimal coupling vertex $\hat{p}_i\cdot\varepsilon\sim (1-a_iz)$.}. 
\end{enumerate}
This yields the central result of our formalism, the \textbf{gauged soft recursion relation}:
\eq{\label{eq:master}
\mc{A}_{i_1,\dots,i_l} &= \frac{1}{2\pi \i}\oint_{z=0}\frac{\d z}{z}\frac{\hat{\mc{A}}_{i_1,\dots,i_l}(z)}{F(z)} \\
&= - \sum_I\res_{z=z_I^\pm}\frac{\hat{\mc{A}}_{I\rm{L}}(z)\times\hat{\mc{A}}_{I\rm{R}}(z)}{zF(z)\hat{P}(z)^2} - \sum_s\res_{z=\frac{1}{a_{s}}} \frac{\hat{\mathcal{A}}_{i_1,\dots,i_l}(z)}{zF(z)}\ .
}
The residues on the hard poles factorize as usual, due to the principle of unitarity, into lower-point amplitudes.
The $l$ gauge bosons are partitioned between $\mc{A}_{I{\rm L/R}}$ according to the charge flows they are associated with. 
The residues on the soft poles, on the other hand, can be constructed from the soft photon theorem as in eq.~\eqref{eq:soft_photon_recursion}
\eq{
    \res_{z=\frac{1}{a_{s}}} \frac{\hat{\mathcal{A}}_{i_1,\dots,i_l}(z)}{zF(z)} = \res_{z=\frac{1}{a_{s}}} 
    \left[
    \frac{\hat{S}^{(0)}\hat{\mathcal{A}}_{i_1,\dots,\slashed{s},\dots,i_l}(z)}{zF(z)(1-a_s z)} + \frac{\hat{S}^{(1)}\hat{\mathcal{A}}_{i_1,\dots,\slashed{s},\dots,i_l}(z)}{zF(z)} 
    \right]
}
where $\mathcal{A}_{i_1,\dots,\slashed{s},\dots,i_l}$ has one less gauge boson than the original amplitude and is regular at $z=1/a_s$. 
Therefore, a generic amplitude component can be constructed recursively from the basic amplitudes with neither hard poles nor soft poles, which are the soft blocks of the Goldstone bosons that satisfy the Adler's zero condition on their own.
In the following we provide examples to demonstrate the gauged soft recursion relations.

%%%%%%%%%%%%%%%%%%%%%%%%%%%%%%%%%%%%%%%%%%%%%%%%%%%%%%%%%%%%%%%%%

\subsection{4 Scalars + 1 Photon}
\label{sec:4+1}

We start from the $4\phi+1\gamma$ (4+1 for short) amplitude $\mathcal{M}_{4+1}(\phi_1^{q_1},\phi_2^{q_2},\phi_3^{q_3},\phi_4^{q_4};{\gamma})$ as an example.
The full amplitude $\mc{M}_{4+1}$ consists of the contribution with internal photon $\mc{M}_{e^3}$, which was discussed in section~\ref{sec:int_ph}, and a contribution $\mc{M}_{e}$ from eq.~\eqref{eq:phi4}.
The latter is decomposed via eqs.~(\ref{eq:sub_amp_ex1q}-\ref{eq:sub_amp_ex1}) as
\begin{align}
    & \mathcal{M}_{e}(\phi_1^{q_1},\phi_2^{q_2},\phi_3^{q_3},\phi_4^{q_4};{\gamma}) = Q_1\mc{A}_1 + Q_2\mc{A}_2 + Q_3\mc{A}_3 \label{eq:phi4charges} \\
    & \left\{\begin{array}{lr}
         Q_1 = q_1\ ,& \mc{A}_1 = \mathcal{M}_{e}(\phi_1^+,\phi_2^-,\phi_3,\phi_4;{\gamma})\ , \\
         Q_2 = q_1+q_2\ ,& \mc{A}_2 = \mathcal{M}_{e}(\phi_1,\phi_2^+,\phi_3^-,\phi_4;{\gamma})\ ,\\
         Q_3 = q_1+q_2+q_3 \ ,& \mc{A}_3 = \mathcal{M}_{e}(\phi_1,\phi_2,\phi_3^+,\phi_4^-;{\gamma})\ .
    \end{array}\right.
\end{align}
Then we proceed to compute the components $\mc{A}_i$ separately. Each of them has two charged scalars and two ``neutral'' scalars, while the latter should satisfy the Adler's Zero condition, for example
\eq{
    \lim_{p_3\to0} \mc{A}_1 = \lim_{p_4\to0} \mc{A}_1 = 0\ .
}

The 4+1 amplitude is a special case because the all line shift $p_i \to (1-a_i)p_i$ is not viable for $n\le D+1$ particles, with $D=4$ the spacetime dimension. The only solution to the momentum conservation and the on-shell conditions would be the trivial case where $a_i = a$ for all particles, which cannot probe the single soft photon pole $p_s^2=0$. In this case, we apply the Risager-type momentum shift \cite{Risager:2005vk}:
we shift only the two charged scalars $(i,j)$ and the gauge boson with positive helicity  as follows
\begin{align}
    \hat{p}_{i,j}=p_{i.j}-za_{i,j}r_{i,j},\quad \hat{p}_s=(1-a_sz)p_s,\quad\text{while }r_{i}=|i\rangle[s|,\;r_{j}=|j\rangle[s|. \label{eq:4+1shift}
\end{align}
We can find the on-shell condition is satisfied automatically, and the momentum conservation requires that
\begin{align}
    a_ir_i+a_jr_j+a_sp_s=0\ ,
\end{align}
which is satisfied by the following solution due to the Schouten Identity
\eq{
    a_i=\vev{js},\;a_j=\vev{si},\;a_s=\vev{ij}\ .
}
The only pole on the $z$-plane would be the soft pole $z=1/a_s$. The amplitude can thus be written as
\eq{\label{eq:A1}
    \mc{A}_1 &= \frac{1}{2\pi\i}\oint_{z=0}\frac{\hat{\mc{A}}_1(z)}{z(1-a_s z)} = -\res_{z=1/a_s}\frac{\hat{\mc{A}}_1(z)}{z(1-a_s z)} \\
    &=-\res_{z=1/a_s}\left( \frac{1}{z(1-a_sz)^2}S^{(0)}\hat{\mc{M}}_4(z) + \frac{1}{z(1-a_sz)}S^{(1)}\hat{\mc{M}}_4(z) \right) \ ,
}
%%%%%%
We can use Cauchy's theorem again to go back to the $z=0$ residue, which simply proves that the amplitude can be given by the first two terms in the soft expansion without higher order correction\footnote{This is common that the soft expansion terminates at finite order when the hard amplitude is local with a finite power of momenta, as the higher order terms in the expansion require higher order derivatives of momenta.}
\eq{\label{eq:phi4ST}
    \mc{A}_1 = \Big(S^{(0)} + S^{(1)}\Big)\mc{M}_4 \ .
}
Assuming $\mc{M}_4 = s_{12}/f^2$, we have
\eq{
    \mc{A}_1 = \frac{e}{f^2}\frac{\vev{12}}{\vev{s1}\vev{s2}}(s_{12} + s_{2s} + s_{1s}) = \frac{e}{f^2}\frac{\vev{12}}{\vev{s1}\vev{s2}}s_{34}\ .
} 
Similarly, we can obtain the other amplitude components
\eq{
    \mc{A}_2 &= \frac{e}{f^2}\left(\frac{\vev{23}}{\vev{s2}\vev{s3}}s_{12} + \frac{\vev{12}[s1]}{\vev{s2}}\right) = \frac{e}{f^2}\frac{\vev{12}\vev{34}[41]}{\vev{s2}\vev{s3}} \ , \\
    \mc{A}_3 &= \frac{e}{f^2}\frac{\vev{34}}{\vev{s3}\vev{s4}}s_{12} \ .
}
Plug them in eq.~\eqref{eq:phi4charges} and one could easily get the contribution $\mc{M}_{e}$ for arbitrary charges $\{q_i\}$. For arbitrary charges of the scalars, one could simply combine the above three components according to eq.~\eqref{eq:phi4charges}, for example
\eq{\label{eq:M4+1}
    \mc{M}(\pi_1^+,\pi^-_2,\pi_3^0,\pi^0_4;\gamma_5)&=\mc{A}_1=\frac{e}{f^2}\frac{\vev{12}s_{34}}{\vev{15}\vev{25}}\ , \\
    \mc{M}(\pi_1^+,\pi^+_2,\pi_3^-,\pi^-_4;\gamma_5)&=-\mc{A}_1-2\mc{A}_2-\mc{A}_3=-\frac{e}{f^2}\frac{\vev{12}\vev{34}[24]}{\vev{15}\vev{35}}-\frac{e}{f^2}\frac{\vev{12}\vev{34}[13]}{\vev{25}\vev{45}}\ .
}

%=======================================================================================================
\subsection{$n$ Scalars + 1 Photon}

Next we can move on to the higher-point amplitudes with hard poles. 
The 6+1 amplitude can be decomposed in the similar way as in eq.~\eqref{eq:phi4charges}. 
\eq{\label{eq:phi6charges}
    & \mc{M}_{6+1}(\pi_1^+,\pi_2^+,\pi_3^0,\pi_4^0,\pi_5^-,\pi_6^-;\gamma) = \mc{A}_1 + \mc{A}_2\ , \\
    & \mc{A}_1 = \mc{M}_{6+1}(\phi_1^+,\phi_2^0,\phi_3^0,\phi_4^0,\phi_5^-,\phi_6^0;\gamma) \ ,\\
    & \mc{A}_2 = \mc{M}_{6+1}(\phi_1^0,\phi_2^+,\phi_3^0,\phi_4^0,\phi_5^0,\phi_6^-;\gamma) \ .
}
We now apply the all-line momentum shift. It is important to note that the sets of neutral scalars differ between $\mc{A}_1$ and $\mc{A}_2$; therefore, the rescaling factor $F(z)$ must be defined separately for each component. To compute $\mc{A}_1$, we have
\begin{align}
    & \hat{p}_i=(1-a_iz)p_i,\quad i=1,2,3,4,5,6,7 \ ,\\
    & F_4(z)= (1-a_2z)(1-a_3z)(1-a_4z)(1-a_6z) \ .
\end{align}
Note that the shifted amplitude behaves as $\lim_{z\to\infty}\hat{\mc{M}}(z) \sim z^1$, thus the 4th power of $z$ in the denominator $F_4(z)$ is more than sufficient to eliminate the boundary term $B_\infty$ in the recursion relation.
The calculation is simpler without another factor $1-a_7 z$ from the gauge boson in that the residue on the soft pole would only involve the $S^{(0)}$ term.
According to eq.~\eqref{eq:master}, the amplitude components can be constructed from the residues on the hard poles and the soft pole, such as
\eq{
    \mc{A}_1 = -\sum_I \res_{z=z_I^\pm}\frac{\hat{\mc{A}}_{I\rm{L}}(z)\times \hat{\mc{A}}_{I\rm{R}}(z)}{zF_4(z)\hat{P}_I^2(z)} - \res_{z=\frac1{a_7}}\frac{\hat{\mc{A}}_1(z)}{zF_4(z)}\ .
}
By applying Cauchy's theorem again for the first term, we obtain
\begin{align}
    \mc{A}_1&=\sum_I\left(\res_{z=0}\frac{\hat{\mc{A}}_{I{\rm L}}(z)\hat{\mc{A}}_{I{\rm R}}(z)}{zF_4(z)\hat{P}_I^2(z)} + \sum_{i}\res_{z=\frac{1}{a_i}}\frac{\hat{\mc{A}}_{I{\rm L}}(z)\hat{\mc{A}}_{I{\rm R}}(z)}{zF_4(z)\hat{P}_I^2(z)} \right)-\res_{z=\frac{1}{a_7}}\frac{\hat{S}^{(0)}\hat{\mc{M}}_6(1^+,\hat{2},\hat{3},\hat{4},5^-,\hat{6})}{zF_4(z)} \notag\\
    &= \sum_I\frac{\mc{A}_{I{\rm L}}\mc{A}_{I{\rm R}}}{P_I^2} + \left(\sum_{i}\res_{z=\frac{1}{a_i}}\sum_I\frac{\hat{\mc{A}}_{I{\rm L}}(z)\hat{\mc{A}}_{I{\rm R}}(z)}{zF_4(z)\hat{P}_I^2(z)} -\res_{z=\frac{1}{a_7}}\frac{S^{(0)}\hat{\mc{M}}_6(1^+,\hat{2},\hat{3},\hat{4},5^-,\hat{6})}{zF_4(z)(1-a_7z)} \right) \label{eq:M6+1RR}\\
    &\equiv \frac{e}{f^4}\left(\mc{A}_1^{(1)}+\mc{A}_2^{(2)} \right) \ , \notag
\end{align}
where for each channel $I$, one of the factors $\mc{A}_{I{\rm L/R}}$ is a 4-point soft block $\mc{M}_4$ and the other being $\mc{M}_{4+1}$ which contains the photon.
A subtle point is that before applying Cauchy's theorem, the $z$ function for which we compute the residues
\[ \frac{\hat{\mc{A}}_{I{\rm L}}(z)\hat{\mc{A}}_{I{\rm R}}(z)}{zF_4(z)\hat{P}_I^2(z)} \]
should be written only in terms of the external kinematic variables and $z$, without referring to the momentum or the spinors $P$ of the internal propagator, which are only well-defined at the $z=z_I^\pm$ poles. There might be different choices of the function, as we will show in examples, but as long as the residues are computed consistently for the same function, Cauchy's theorem and the constructibility condition $\res_{z=\infty}\cdots=0$ guarantee the uniqueness of the sum in eq.~\eqref{eq:M6+1RR}.
Similar as in eq.~\eqref{eq:M6RR}, the amplitude is divided into two parts, one accounting for the tree-level Feynman diagrams
\eq{
    \mc{A}_1^{(1)} = \sum_I\frac{\mc{A}_{I{\rm L}}\mc{A}_{I{\rm R}}}{P_I^2}\ ,
}
and the other supposed to account for contributions involving higher point corrections in the Goldstone Lagrangian
\eq{\label{eq:A1(2)}
    \mc{A}_1^{(2)} = \sum_{i=2,3,4,6,7}\res_{z=\frac{1}{a_i}}\frac{1}{zF_4(z)} \sum_I\frac{\hat{\mc{A}}_{I{\rm L}}(z)\hat{\mc{A}}_{I{\rm R}}(z)}{\hat{P}_I^2(z)} - \res_{z=\frac{1}{a_7}}\frac{1}{zF_4(z)} \frac{S^{(0)}\hat{\mc{M}}_6(1,\hat{2},\hat{3},\hat{4},5,\hat{6})}{1-a_7 z}\ .
}

Now we compute the two parts based on the known results in eq.~\eqref{eq:pi4_bb},~\eqref{eq:M6RR} and~\eqref{eq:M4+1}. The sum over hard poles can be identified from the same scattering channels as in figure~\ref{fig:pi60}, which are classified into two categories, one with the charged scalars $\phi_1^+,\phi_5^-$ on the same side (remember for $\mc{A}_1$ we are treating $\phi_2$ and $\phi_6$ as neutral), and the other with them on both sides. An example of the former is the first channel $I=(246)$,
\eq{
    \left.\frac{\hat{\mc{A}}_{I{\rm L}}(z)\hat{\mc{A}}_{I{\rm R}}(z)}{\hat{P}_I^2(z)}\right|_{I=(246)} &= \frac{1}{P_{246}^2} \hat{\mc{M}}_{4+1}(\phi_1^+,\phi_3,\phi_5^-,\phi';\gamma_7) \times \hat{\mc{M}}_4(\phi_2,\phi_4,\phi_6,\phi')\ . % \\
    % &= \frac{1}{s_{246}} \left[ \frac{e}{f^2} \frac{\vev{15}}{\vev{17}\vev{57}}(s_{23}+s_{34}+s_{36}) \right] \times \frac{s_{26}}{f^2}
}
The function $\hat{\mc{M}}_4$ is given in eq.~\eqref{eq:pi4_bb} as $\hat{s}_{26}/f^2$, while $\hat{\mc{M}}_{4+1}$ is written based on eq.~\eqref{eq:M4+1} as
\eq{\label{eq:hatM4+1}
    \hat{\mc{M}}_{4+1}(\phi_1^+,\phi_3,\phi_5^-,\phi'(\hat{P});\gamma_7) = \frac{e}{f^2}\frac{\vev{15}\hat{s}_{3P}}{\vev{1\hat{7}}\vev{5\hat{7}}} = \frac{e}{f^2}\frac{\vev{15}}{\vev{1\hat{7}}\vev{5\hat{7}}}(\hat{s}_{23}+\hat{s}_{34}+\hat{s}_{36}) \ .
}
where we used $\hat{P}=\hat{p}_2+\hat{p}_3+\hat{p}_6$. Taking the $z=0$ value, we obtain 
\eq{\label{eq:A11ex}
    \mc{A}_1^{(1)} = \frac{e}{f^4}\frac{\vev{15}}{\vev{17}\vev{57}}\frac{s_{26}}{s_{246}}(s_{23}+s_{34}+s_{36}) \ .
}
As advertized previously, eq.~\eqref{eq:hatM4+1} is not the only way to express $\hat{\mc{M}}_{4+1}$ beyond the hard pole, but we shall evaluate the residues of the same function in $\mc{A}_1^{(2)}$ as
\eq{\label{eq:A12ex1}
    \mc{A}_1^{(2)} &\supset \frac{e}{f^4} \sum_{i=2,3,4,6,7}\res_{z=\frac{1}{a_i}}\frac{1}{zF_4(z)} \frac{\vev{15}}{\vev{1\hat{7}}\vev{5\hat{7}}}\times \frac{\hat{s}_{26}}{\hat{s}_{246}}(\hat{s}_{23}+\hat{s}_{34}+\hat{s}_{36}) \\
    &= \frac{e}{f^4} \frac{\vev{15}}{\vev{17}\vev{57}}\left[ \res_{z=\frac{1}{a_4}}\frac{1}{zF_5(z)} (\hat{s}_{23}+\hat{s}_{36}) 
    + \res_{z=\frac{1}{a_7}}\frac{1}{zF_5(z)}\frac{\hat{s}_{26}}{\hat{s}_{246}}(\hat{s}_{23}+\hat{s}_{34}+\hat{s}_{36})  \right]\ ,
}
where $F_5(z) = F_4(z)\times (1-a_7 z)$. The same channel also contributes to the second term in eq.~\eqref{eq:A1(2)} by the soft operator $S^{(0)}$ acting on the corresponding term in eq.~\eqref{eq:M6RR1}
\eq{\label{eq:A12ex2}
    \mc{A}_1^{(2)} &\supset - \frac{e\vev{15}}{\vev{17}\vev{57}} \res_{z=\frac{1}{a_7}}\frac{1}{zF_5(z)} \left(\frac{1}{f^4}\frac{\hat{s}_{15}\hat{s}_{26}}{\hat{s}_{246}}\right) \ .
}
Adding eq.~\eqref{eq:A11ex}~\eqref{eq:A12ex1} and~\eqref{eq:A12ex2} together, we get the contribution from the first channel in figure~\ref{fig:pi60} as
\eq{
    \mc{A}_1 &\supset \frac{e}{f^4} \frac{\vev{15}}{\vev{17}\vev{57}}\left[ \frac{s_{26}}{s_{246}}(s_{23}+s_{34}+s_{36}) + \res_{z=\frac{1}{a_4}}\frac{1}{zF_5(z)} (\hat{s}_{23}+\hat{s}_{36}) 
    - \res_{z=\frac{1}{a_7}}\frac{1}{zF_5(z)}\hat{s}_{26}  \right]\ .
}
\begin{figure}[htbp]
    \centering

\tikzset{every picture/.style={line width=0.75pt}} %set default line width to 0.75pt        

\begin{tikzpicture}[x=0.75pt,y=0.75pt,yscale=-1,xscale=1]
%uncomment if require: \path (0,161); %set diagram left start at 0, and has height of 161

%Curve Lines [id:da27138431818421327] 
\draw    (411.77,77.03) .. controls (426.66,73.76) and (413.35,95.15) .. (428.24,91.88) ;
%Curve Lines [id:da1950533994047816] 
\draw    (77.94,91.11) .. controls (74.76,76.2) and (96.07,89.65) .. (92.88,74.74) ;
%Straight Lines [id:da7451937882814891] 
\draw    (58.1,71.82) -- (190.9,72) ;
%Straight Lines [id:da11444738239477092] 
\draw    (92.88,44.65) -- (92.88,104.83) ;
%Straight Lines [id:da0623067450738336] 
\draw    (156.12,44.65) -- (156.12,104.83) ;
%Straight Lines [id:da712680893526959] 
\draw [color={rgb, 255:red, 74; green, 144; blue, 226 }  ,draw opacity=1 ] [dash pattern={on 4.5pt off 4.5pt}]  (124.5,20.86) -- (124.5,122.97) ;
%Straight Lines [id:da5378082294585939] 
\draw    (313.1,71.82) -- (445.9,72) ;
%Straight Lines [id:da4970748421414172] 
\draw    (347.88,44.65) -- (347.88,104.83) ;
%Straight Lines [id:da5470721127808363] 
\draw    (411.12,44.65) -- (411.12,104.83) ;
%Straight Lines [id:da0507690655175943] 
\draw [color={rgb, 255:red, 74; green, 144; blue, 226 }  ,draw opacity=1 ] [dash pattern={on 4.5pt off 4.5pt}]  (379.5,20.86) -- (379.5,122.97) ;
%Shape: Circle [id:dp2606922341455299] 
\draw  [fill={rgb, 255:red, 74; green, 144; blue, 226 }  ,fill opacity=1 ] (79.18,74.74) .. controls (79.18,67.17) and (85.32,61.04) .. (92.88,61.04) .. controls (100.45,61.04) and (106.59,67.17) .. (106.59,74.74) .. controls (106.59,82.31) and (100.45,88.44) .. (92.88,88.44) .. controls (85.32,88.44) and (79.18,82.31) .. (79.18,74.74) -- cycle ;
%Shape: Circle [id:dp8159306037260707] 
\draw  [fill={rgb, 255:red, 74; green, 144; blue, 226 }  ,fill opacity=1 ] (397.42,74.74) .. controls (397.42,67.17) and (403.55,61.04) .. (411.12,61.04) .. controls (418.69,61.04) and (424.82,67.17) .. (424.82,74.74) .. controls (424.82,82.31) and (418.69,88.44) .. (411.12,88.44) .. controls (403.55,88.44) and (397.42,82.31) .. (397.42,74.74) -- cycle ;
%Curve Lines [id:da8198805232926234] 
\draw    (63,107.48) .. controls (59.81,92.57) and (81.12,106.02) .. (77.94,91.11) ;
%Curve Lines [id:da819423952095295] 
\draw    (48.05,123.86) .. controls (44.87,108.95) and (66.18,122.39) .. (63,107.48) ;
%Curve Lines [id:da23436530948614553] 
\draw    (428.24,91.88) .. controls (443.13,88.6) and (429.81,109.99) .. (444.7,106.72) ;
%Curve Lines [id:da8667536461644797] 
\draw    (444.7,106.72) .. controls (459.59,103.45) and (446.27,124.84) .. (461.16,121.56) ;
%Straight Lines [id:da4167952512603865] 
\draw [color={rgb, 255:red, 80; green, 227; blue, 194 }  ,draw opacity=1 ]   (109.63,80.58) -- (151.5,80) ;
\draw [shift={(135.56,80.22)}, rotate = 179.21] [fill={rgb, 255:red, 80; green, 227; blue, 194 }  ,fill opacity=1 ][line width=0.08]  [draw opacity=0] (10.72,-5.15) -- (0,0) -- (10.72,5.15) -- (7.12,0) -- cycle    ;
%Straight Lines [id:da9515493414191197] 
\draw [color={rgb, 255:red, 80; green, 227; blue, 194 }  ,draw opacity=1 ]   (151.5,56) -- (151.5,80) ;
%Straight Lines [id:da5945284894953987] 
\draw [color={rgb, 255:red, 80; green, 227; blue, 194 }  ,draw opacity=1 ]   (109.63,80.58) -- (108.63,111.58) ;
%Straight Lines [id:da9422987484646063] 
\draw [color={rgb, 255:red, 80; green, 227; blue, 194 }  ,draw opacity=1 ]   (354.63,64.58) -- (396.5,64) ;
\draw [shift={(380.56,64.22)}, rotate = 179.21] [fill={rgb, 255:red, 80; green, 227; blue, 194 }  ,fill opacity=1 ][line width=0.08]  [draw opacity=0] (10.72,-5.15) -- (0,0) -- (10.72,5.15) -- (7.12,0) -- cycle    ;
%Straight Lines [id:da254727938046024] 
\draw [color={rgb, 255:red, 80; green, 227; blue, 194 }  ,draw opacity=1 ]   (396.5,40) -- (396.5,64) ;
%Straight Lines [id:da4138723378531276] 
\draw [color={rgb, 255:red, 80; green, 227; blue, 194 }  ,draw opacity=1 ]   (354.63,64.58) -- (353.63,95.58) ;

% Text Node
\draw (86.05,105.89) node [anchor=north west][inner sep=0.75pt]    {$\pi _{1}^{+}$};
% Text Node
\draw (147.7,105.89) node [anchor=north west][inner sep=0.75pt]    {$\pi _{2}^{+}$};
% Text Node
\draw (39.09,67.6) node [anchor=north west][inner sep=0.75pt]    {$\pi _{3}^{0}$};
% Text Node
\draw (195.08,66.38) node [anchor=north west][inner sep=0.75pt]    {$\pi _{4}^{0}$};
% Text Node
\draw (88.15,25.66) node [anchor=north west][inner sep=0.75pt]    {$\pi _{6}^{-}$};
% Text Node
\draw (150.87,25.66) node [anchor=north west][inner sep=0.75pt]    {$\pi _{5}^{-}$};
% Text Node
\draw (341.05,105.89) node [anchor=north west][inner sep=0.75pt]    {$\pi _{1}^{+}$};
% Text Node
\draw (402.7,105.89) node [anchor=north west][inner sep=0.75pt]    {$\pi _{2}^{+}$};
% Text Node
\draw (294.09,67.6) node [anchor=north west][inner sep=0.75pt]    {$\pi _{3}^{0}$};
% Text Node
\draw (450.08,66.38) node [anchor=north west][inner sep=0.75pt]    {$\pi _{4}^{0}$};
% Text Node
\draw (343.15,25.66) node [anchor=north west][inner sep=0.75pt]    {$\pi _{6}^{-}$};
% Text Node
\draw (405.87,25.66) node [anchor=north west][inner sep=0.75pt]    {$\pi _{5}^{-}$};
% Text Node
\draw (28,119.4) node [anchor=north west][inner sep=0.75pt]    {$\gamma _{7}$};
% Text Node
\draw (465,119.4) node [anchor=north west][inner sep=0.75pt]    {$\gamma _{7}$};

\end{tikzpicture}

    \caption{Here are the $\mathcal{M}(\pi_1^+,\pi_2^+,\pi_3^0,\pi_4^0,\pi_5^-,\pi_6^-;\gamma)$ diagrams for the channels $I=(136)$ and $I=(245)$. In the $\mc{A}_1$ component, only the particles 1 and 5 carry charges. The effective current is indicated by the light green line with arrow. The current flows through the channel, making the intermediate state effectively charged in this component, although it is a neutral state in the full amplitude.
    }
    \label{fig:pi61}
\end{figure}
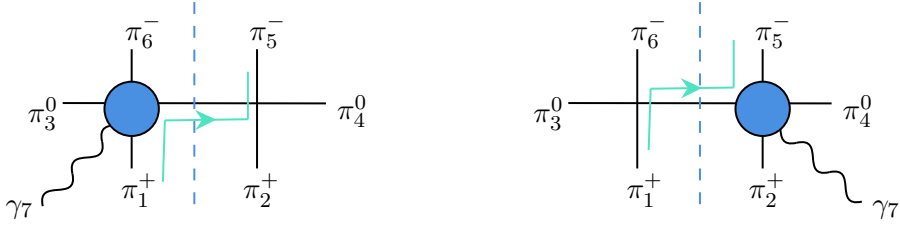

Let's look at another example of poles $I=(136)$ and $I=(245)$ as shown in figure~\ref{fig:pi61},  which originate from the 3rd diagram in figure~\ref{fig:pi60} where the photon could attach to either side
\eq{\label{eq:6+1_f3}
    % \mc{A}_1^{(1)} &\supset 
    \sum_{I}\frac{\hat{\mc{A}}_{I{\rm L}}(z)\hat{\mc{A}}_{I{\rm R}}(z)}{\hat{P}_I^2(z)} &\supset \frac{1}{\hat{s}_{136}} \hat{\mc{M}}_{4}(\phi_1^+,\phi_3,\phi_6,\phi') \times \hat{\mc{M}}_{4+1}(\phi_2,\phi_4,\phi_5^-,\phi'{}^*;\gamma_7) \\
    & \qquad\qquad + \frac{1}{\hat{s}_{245}} \hat{\mc{M}}_{4+1}(\phi_1^+,\phi_3,\phi_6,\phi';\gamma_7) \times \hat{\mc{M}}_{4}(\phi_2,\phi_4,\phi_5^-,\phi'{}^*) \ .
}
The subamplitudes $\mc{M}_{4+1}$ involve the photon $\gamma_7$ which recognizes the charge currents designated for the amplitude component $\mc{A}_1$, with $\phi_2$ and $\phi_6$ neutral and the intermediate particle $\phi'$ negatively charged. Note that $\phi'$ in the same channel would be positively charged in $\mc{A}_2$, and it is actually neutral in the full amplitude after the summation. On the other hand, the 4-point amplitude $\mc{M}_4$, upon which $\mc{M}_{4+1}$ is built, should be defined via the original scalar flavors as
\eq{
    \mc{M}_4(\phi_1^+,\phi_3,\phi_6,\phi') & \equiv \mc{M}_4(\pi^+,\pi^0,\pi^-,\pi^0) = \frac{s_{16}}{f^2}\ ,\\ 
    \mc{M}_4(\phi_2,\phi_4,\phi_5^-,\phi'{}^*) & \equiv \mc{M}_4(\pi^+,\pi^0,\pi^-,\pi^0) = \frac{s_{25}}{f^2} \ .
}
We showed in the section~\ref{sec:4+1} that the $\mc{M}_{4+1}$ can be derived solely by the soft photon theorem
\eq{ 
    \hat{\mc{M}}_{4+1}(\phi_2,\phi_4,\phi_5^-,\phi'{}^*(\hat{P});\gamma_7^{(+)}) &= (\hat{S}^{(0)} + \hat{S}^{(1)})\hat{\mc{M}}_4(\phi_2,\phi_4,\phi_5^-,\phi'{}^*(\hat{P})) \\
    &=e\left(\frac{\vev{\hat{P}5}}{\vev{\hat{P}\hat{7}}\vev{5\hat{7}}}+\frac{[\hat{7}|}{\vev{\hat{7}\hat{P}}}\frac{\partial}{\partial [\hat{P}|}-\frac{[\hat{7}|}{\vev{\hat{7}5}}\frac{\partial}{\partial [5|}\right)\frac{\hat{s}_{25}}{f^2} \\
    &= \frac{e}{f^2}\left(\frac{ [\hat{7}|\hat{P}\ket{5} }{[\hat{7}|\hat{P}\ket{7}\vev{5\hat{7}}}\hat{s}_{25} + \frac{[\hat{7}\hat{2}]\vev{\hat{2}5}}{\vev{5\hat{7}}} \right) \\
    &= -\frac{e}{f^2}\left(\frac{\hat{s}_{25}}{\hat{s}_{245}-\hat{s}_{136}}\sum_{2,4}\frac{ [\hat{7}\hat{i}]\vev{\hat{i}5} }{\vev{5\hat{7}}} - \frac{[\hat{7}\hat{2}]\vev{\hat{2}5}}{\vev{5\hat{7}}} \right) \ .
}
In the last line, we express the result replacing the internal momentum $\hat{P}$, defined only at the hard pole, by momentum conservation $\hat{P}=\hat{p}_1+\hat{p}_3+\hat{p}_6=-\hat{p}_2-\hat{p}_4-\hat{p}_5-\hat{p}_7$. Similar computation can be done for the $I=(245)$ pole
\eq{
    \hat{\mc{M}}_{4+1}(\phi_1^+,\phi_3,\phi_6,\phi'{}^*(\hat{P});\gamma_7^{(+)}) = \frac{e}{f^2}\left( \frac{\hat{s}_{16}}{\hat{s}_{136}-\hat{s}_{245}}\sum_{i=3,6}\frac{[\hat{7}\hat{i}]\vev{\hat{i}1}}{\vev{1\hat{7}}} - \frac{[\hat{7}\hat{6}]\vev{\hat{6}1}}{\vev{1\hat{7}}} \right)\ .
}
Now eq.~\eqref{eq:6+1_f3} can be evaluated as
% Therefore, we end up with $\mc{A}_1^{(1)}$ as the residue at the $z=0$ pole as
\eq{
    % \mc{A}_1^{(1)} &\supset 
    & \frac{e}{f^4} \left[\frac{\<15\>s_{16}s_{25}}{\<17\>\<57\>s_{245}} -\sum_{i=2,4}\frac{[7i]\<i5\>s_{16}s_{25}}{\<57\>s_{136}s_{245}} -\frac{\<16\>[67]}{\<17\>}\frac{s_{25}}{s_{245}}+\frac{\<52\>[27]}{\<57\>}\frac{s_{16}}{s_{136}} \right] \\
   =& \frac{e}{f^4} \left[\frac{\<15\>s_{16}s_{25}}{\<17\>\<57\>s_{136}} +\sum_{i=3,6}\frac{[7i]\<i1\>s_{16}s_{25}}{\<17\>s_{136}s_{245}} -\frac{\<16\>[67]}{\<17\>}\frac{s_{25}}{s_{245}}+\frac{\<52\>[27]}{\<57\>}\frac{s_{16}}{s_{136}} \right].
}
whose unhatted version is part of $\mc{A}_1^{(1)}$ and its residues at the $1/a_i$ poles contribute to $\mc{A}_1^{(2)}$
\eq{
    \mc{A}_1^{(2)} &\supset 
    \res_{z=\frac{1}{a_3}} \frac{1}{zF_4(z)}\left(\frac{\vev{15}}{\vev{1\hat{7}}\vev{5\hat{7}}}\hat{s}_{25} + \frac{\vev{5\hat{2}}[\hat{2}\hat{7}]}{\vev{5\hat{7}}} \right) + \res_{z=\frac{1}{a_4}} \frac{1}{zF_4(z)}\left( \frac{\vev{15}}{\vev{1\hat{7}}\vev{5\hat{7}}}\hat{s}_{16} - \frac{\vev{1\hat{6}}[\hat{6}\hat{7}]}{\vev{1\hat{7}}}\right) \ .
}
Note that the $1/a_7$ residue is cancelled by the second term in eq.~\eqref{eq:A1(2)}.

These residues need not be explicitly computed. Just like those in the original soft recursion relation eq.~\eqref{eq:M6RR2}, after adding up all the factorization channels, they can always be written as a sum over the $1/a_i$ residues of a common function of $z$, and then turn to the $z=0$ residue by another application of Cauchy's theorem. In this case, the final result is

\eq{
    \mc{A}_1^{(1)} &= \left(\frac{\<15\>}{\<17\>\<57\>}(s_{32}+s_{34}+s_{36})\frac{s_{26}}{s_{246}}+3\leftrightarrow4\right)\\
    &\quad +\left(\frac{\<15\>s_{16}s_{25}}{\<17\>\<57\>s_{245}} -\sum_{i=2,4}\frac{[7i]\<i5\>s_{16}s_{25}}{\<57\>s_{136}s_{245}} -\frac{\<16\>[67]}{\<17\>}\frac{s_{25}}{s_{245}}+\frac{\<52\>[27]}{\<57\>}\frac{s_{16}}{s_{136}} +3\leftrightarrow4\right)\\
    &\quad +\sum_{i=5,7}\frac{\<1i\>\<56\>[i6]}{\<17\>\<57\>}\frac{s_{34}}{s_{234}}
    -\frac{\<56\>[67]s_{34}}{\<57\>s_{134}}-\frac{\<15\>s_{34}s_{56}}{\<17\>\<57\>s_{256}}+\sum_{i=2,6}\frac{[7i]\<i5\>s_{34}s_{56}}{\<57\>s_{134}s_{256}}\\
    &\quad +\sum_{i=1,7}\frac{\<12\>\<5i\>[2i]}{\<17\>\<57\>}\frac{s_{34}}{s_{346}}
    +\frac{\<12\>[27]s_{34}}{\<17\>s_{345}}-\frac{\<15\>s_{12}s_{34}}{\<17\>\<57\>s_{126}}-\sum_{i=2,6}\frac{[7i]\<i1\>s_{12}s_{34}}{\<17\>s_{126}s_{345}}\ ,
}
\eq{
    \mc{A}_1^{(2)} &= \sum_{i=2,3,4,6,7}\res_{z=\frac{1}{a_i}}\frac{1}{zF_4(z)}\left(-\frac{\<15\>}{\<1\hat{7}\>\<5\hat{7}\>}\left(2\hat{s}_{26}+\hat{s}_{12}+\hat{s}_{34}+\hat{s}_{56}\right)
    -\frac{\<5\hat{6}\>[\hat{6}7]}{\<57\>}
    +\frac{\<1\hat{2}\>[\hat{2}7]}{\<17\>}\right)\\
    &=\frac{\<15\>}{\<17\>\<57\>}\left(2s_{26}+s_{12}+s_{34}+s_{56}\right)
    +\frac{\<56\>[67]}{\<57\>}
    -\frac{\<12\>[27]}{\<17\>}\ .
}

\input{Fig_phi6A}

We present another example for the fundamental representation of $SU(N)\times U(1)$, in which the pure Goldstone building blocks are shown in eq.~\eqref{eq:SUN4p} and eq.~\eqref{eq:SUN6p}. First, we define the partial amplitude $\mc{A}$ as
\begin{align}
    \mathcal{M}_{6+1}=\delta^{i_1}_{i_4}\delta^{i_2}_{i_5}\delta^{i_3}_{i_6}\mc{A}+\text{permutations of }(4,5,6) \ ,
\end{align}
which no longer carries the group indices. Assuming $\phi^i$ has $+1$ charge under the $U(1)$ gauge group, its conjugate $\bar\phi_i$ would carry $-1$ charge, so we have all six scalars carrying non-zero charges. We have to do a charge decomposition as in eq.~\eqref{eq:phi6charges} before the soft recursion, one of the choice being
\eq{
    \mc{A}\big(\phi_1^{+1},&\phi_2^{+1},\phi_3^{+1},\phi_4^{-1},\phi_5^{-1},\phi_6^{-1};\gamma_7\big)=\mathcal{A}_1+\mathcal{A}_2+\mathcal{A}_3 \ ,\\
    &\mathcal{A}_1=\mc{A}\left(\phi_1^{+1},\phi_2^{0},\phi_3^{0},\phi_4^{-1},\phi_5^{0},\phi_6^{0};\gamma_7\right)\ ,\\
    &\mathcal{A}_2=\mc{A}\left(\phi_1^{0},\phi_2^{+1},\phi_3^{0},\phi_4^{0},\phi_5^{-1},\phi_6^{0};\gamma_7\right)\ ,\\
    &\mathcal{A}_3=\mc{A}\left(\phi_1^{0},\phi_2^{0},\phi_3^{+1},\phi_4^{0},\phi_5^{0},\phi_6^{-1};\gamma_7\right)\ .
}
It is easy to recognize that $\mc{A}_{2,3}$ are just permutations of $\mc{A}_1$, while $\mc{A}_1$ has the factorization channels depicted in figure~\ref{fig:phi6A}.
The necessary building block $\mc{M}_{4+1}$ follows from the same shift and calculation as in eq.~\eqref{eq:4+1shift} and eq.~\eqref{eq:A1}:
\begin{align}
\begin{split}
    \mathcal{M}_{4+1}\left((\phi_1^+)^{i_1},(\phi_2^0)^{i_2},(\bar{\phi}_3^-)_{i_3},(\bar{\phi}_4^0)_{i_4};\gamma_5^{(+)}\right)=\delta^{i_1}_{i_3}\delta^{i_2}_{i_4}\frac{e}{f^2}\frac{\vev{13}s_{24}}{\vev{15}\vev{35}}+\delta^{i_1}_{i_4}\delta^{i_2}_{i_3}\frac{e}{f^2}\frac{\vev{14}\vev{23}[24]}{\vev{15}\vev{35}}\ .
\end{split}
\end{align}
Under the all-line shift $\hat{p}_i=p_i(1-a_iz)$, the amplitude $\mc{A}_1$ can be obtained from the hard and soft poles as a sum of the following two contributions
\begin{align}
    \mc{A}_1^{(1)}
    &=\frac{s_{25}s_{36}}{s_{125}s_{346}}\frac{\sum_{2,5}[7i]\langle i1\rangle}{\langle 17\rangle}+ \frac{s_{25}s_{36}}{s_{136}s_{245}}\frac{\sum_{3,6}[7i]\langle i1\rangle}{\langle 17\rangle} \notag\\
    &\quad + \frac{\langle 14\rangle}{\langle 17\rangle\langle 47\rangle}\Bigg(\frac{s_{25}s_{36}}{s_{125}}+ \frac{s_{25}s_{36}}{s_{136}} +\frac{s_{36}}{s_{356}}(s_{23}+s_{25}+s_{26})+\frac{s_{25}}{s_{256}}(s_{23}+s_{35}+s_{36}) \notag\\
    &\quad\quad +\frac{s_{36}}{s_{236}}(s_{25}+s_{35}+s_{56}) +\frac{s_{25}}{s_{235}}(s_{26}+s_{36}+s_{56}) \Bigg)\ , \\
    \mc{A}_1^{(2)}&=-\frac{\langle 14\rangle}{\langle 17\rangle\langle 47\rangle}\left(s_{23}+s_{26}+s_{35}+s_{56}\right)\ .
\end{align}

\subsection{Two and More Photons}
Next, we apply the gauged soft recursion to the scattering amplitude including more than one gauge boson. 
We take the $\mc{M}_{4+2}(\phi_1,\phi_2,\phi_3,\phi_4;\gamma_5,\gamma_6)$ as a typical example, which is a new building block in the factorizations of $\mc{M}_{n+2}$. Since there are $6>D+1$ particles, we can make the all-line shift of the momenta, so that the amplitude scales as $\lim_{z\to\infty}\mc{M}_{4+2}\sim z^0$. Therefore
\eq{
    \mc{M}_{4+2}(\phi_1,\phi_2,\phi_3,\phi_4;\gamma_5,\gamma_6) = -\left(\res_{z=1/a_5} + \res_{z=1/a_6} + \sum_I\res_{z=z_I^\pm}\right)\frac{\hat{\mc{M}}_{4+2}(z)}{zF_2(z)}\ ,
}
with $F_2(z) = (1-a_5z)(1-a_6z)$. It turns out that the amplitude with photons of the same helicity and that with photons of opposite helicities are qualitatively different. For the same-helicity amplitude, the hard poles have vanishing residues due to $\mc{M}(\phi,\gamma^{(+)},\phi,\gamma^{(+)})=0$ \cite{Elvang:2013cua}, thus we have
\eq{
    \mc{M}^{(s)}_{4+2}(\phi_1,\phi_2,\phi_3,\phi_4;\gamma_5^{(\pm)},\gamma_6^{(\pm)}) = -\res_{z=1/a_5}\frac{(\hat{S}^{(0)}_5 + \hat{S}^{(1)}_5)\hat{\mc{M}}_{4+(6)}}{zF_2(z)} -\res_{z=1/a_6}\frac{(\hat{S}^{(0)}_6 + \hat{S}^{(1)}_6)\hat{\mc{M}}_{4+(5)}}{zF_2(z)} \ .
}
The $\mc{M}_{4+1}$ was proved to be given by the soft photon theorem as
\eq{
    \mc{M}_{4+(a)} = (S^{(0)}_a + S^{(1)}_a)\mc{M}_4 \ ,\quad a=5,6\ .
}
Hence the two terms correspond to the soft operators of $\gamma_5,\gamma_6$ acting on $\mc{M}_4$ with opposite orders. It is easy to show that the soft factors $S^{(0),(1)}$ of photons with same helicities commute with each other, therefore the two terms are the residues of an identical function, for which we may use Cauchy's theorem again
\eq{\label{eq:4+2s}
    \mc{M}^{(s)}_{4+2} &= -\left(\res_{z=1/a_5}+\res_{z=1/a_6}\right)\frac{1}{zF_2(z)}(\hat{S}^{(0)}_5 + \hat{S}^{(1)}_5)(\hat{S}^{(0)}_6 + \hat{S}^{(1)}_6)\hat{\mc{M}}_{4} \\
    &= (S^{(0)}_5 + S^{(1)}_5)(S^{(0)}_6 + S^{(1)}_6)\mc{M}_{4} \ .
}
which simply means that, just like the $\mc{M}_{4+1}$, the amplitude $\mc{M}_{4+2}$ with same-helicity photons can also be given by the leading terms in the soft photon expansion.

The computation of the opposite-helicity amplitude $\mc{M}_{4+2}(\phi_1,\phi_2,\phi_3,\phi_4;\gamma_5^{(+)},\gamma_6^{(-)})$ is more complicated, since the Compton scattering amplitude is not vanishing
\eq{
    \mc{M}^C(\phi_i^{+q},\phi_j^{-q},\gamma_5^{(+)},\gamma_6^{(-)}) = q^2 \frac{\vev{6i}\vev{6j}}{\vev{5i}\vev{5j}} = q^2 \frac{[5i][5j]}{[6i][6j]} = -q^2\frac{[5|p_i\ket{6}}{[6|p_i\ket{5}} \ ,
}
so that the amplitude has a hard pole on each of the charged legs. Moreover, the soft operators of the opposite-helicity gauge bosons do not commute due to the spinor derivative in $S^{(1)}$. Therefore we have to apply the soft recursion relation with caution:
\eq{\label{eq:4+2o}
    & \mc{M}^{(o)}_{4+2}(\phi_1^{q_1},\phi_2^{q_2},\phi_3^{q_3},\phi_4^{q_4};\gamma_5^{(+)},\gamma_6^{(-)}) = \mc{M}^{(1)} + \mc{M}^{(2)} \ ,\quad \mc{M}^{(1)} = \sum_i\frac{\mc{M}_i^C}{s_{i56}}\mc{M}_4 \\
    & \mc{M}^{(2)} = \res_{z=1/a_5}\frac{1}{zF_2(z)} \left[ \sum_i\frac{\mc{M}_i^C}{\hat{s}_{i56}} - (\hat{S}_5^{(0)}+\hat{S}_5^{(1)})(\hat{S}_6^{(0)}+\hat{S}_6^{(1)})\right]\hat{\mc{M}}_4 \\
    &\hspace{2em} + \res_{z=1/a_6}\frac{1}{zF_2(z)} \left[ \sum_i\frac{\mc{M}_i^C}{\hat{s}_{i56}} - (\hat{S}_6^{(0)}+\hat{S}_6^{(1)})(\hat{S}_5^{(0)}+\hat{S}_5^{(1)})\right]\hat{\mc{M}}_4
}
The soft operators of the two photons contribute 4 terms. The leading order $S^{(0)}$ is multiplicative, so $S^{(0)}_5 S^{(0)}_6 = S^{(0)}_6 S^{(0)}_5$, and the two residues of it can be turned to $S^{(0)}_5 S^{(0)}_6\mc{M}_4$ by Cauchy's theorem. Next, we look at the contribution
\eq{
    - \res_{z=1/a_5}\frac{1}{zF_2(z)} \hat{S}_5^{(0)}\hat{S}_6^{(1)} \hat{\mc{M}}_4 - \res_{z=1/a_6}\frac{1}{zF_2(z)} \hat{S}_6^{(1)}\hat{S}_5^{(0)} \hat{\mc{M}}_4 \ .
}
The two functions differ by the order of the soft factors, for which we have
\eq{
    \hat{S}_6^{(1)}\hat{S}_5^{(0)} = \sum_{i} q_i^2\frac{\vev{6i}}{[6i]\vev{\hat{5}\hat{i}}^2} + \hat{S}_5^{(0)}\hat{S}_6^{(1)}
}
while the first term evaluated at the $z=1/a_6$ pole can be written as
\eq{
    \sum_{i} q_i^2\frac{\vev{6i}}{[6i]\vev{\hat{5}\hat{i}}^2} \Big|_{p_6=0} = \sum_{i} -q_i^2\frac{\vev{6i}[i5]}{[6i]\vev{i5}}\frac{1}{\hat{s}_{i5}} \Big|_{p_6=0} = \sum_i \frac{\mc{M}^C_i}{\hat{s}_{i56}} \Big|_{p_6=0}
}
and cancels the first term in the large square bracket in eq.~\eqref{eq:4+2o}. Hence the residues of $\hat{S}_5^{(0)}\hat{S}_6^{(1)}\hat{\mc{M}}_4$ can be combined and turned to the $z=0$ residue, 
\eq{
    - \res_{z=1/a_5}\frac{1}{zF_2(z)} \hat{S}_5^{(0)}\hat{S}_6^{(1)} \hat{\mc{M}}_4 - \res_{z=1/a_6}\frac{1}{zF_2(z)} \left[ \hat{S}_6^{(1)}\hat{S}_5^{(0)} - \frac{\mc{M}_i^C}{\hat{s}_{i56}} \right] \hat{\mc{M}}_4 = S_5^{(0)}S_6^{(1)} \mc{M}_4\ .
}
and so does the term $\hat{S}_6^{(0)}\hat{S}_5^{(1)}\hat{\mc{M}}_4$. Finally, we have
\eq{\label{eq:4+2o11}
    \hat{S}_5^{(1)}\hat{S}_6^{(1)} &= \sum_{i,j}\frac{q_iq_j}{\vev{5\hat{i}}[6\hat{j}]}[5\hat{\partial}_i]\vev{6\hat{\partial}_i} + \sum_i\frac{q_i^2}{\vev{5\hat{i}}}\left([5\hat{\partial}_i]\frac{1}{[6\hat{i}]}\right)\vev{6\hat{\partial}_i} 
}
Consider the second term acting on $\hat{p}_i = \ket{\hat{i}}[\hat{i}|$, we obtain
\eq{
    \frac{q_i^2}{\vev{5\hat{i}}}\left([5\hat{\partial}_i]\frac{1}{[6\hat{i}]}\right)\vev{6\hat{\partial}_i} \ket{\hat{i}}[\hat{i}| \Big|_{\hat{p}_5=0} 
    &= q_i^2\frac{[56]}{\vev{5\hat{i}}[6\hat{i}]^2}\ket{6}[\hat{i}| \\
    &= \left[-q_i^2\frac{[5\hat{i}]\vev{\hat{i}6}}{\vev{5\hat{i}}[\hat{i}6]}\times\frac{\ket{6}[6|}{\hat{s}_{i6}} - q_i^2\frac{\ket{6}[5|}{\vev{5\hat{i}}[6\hat{i}]} \right]\Big|_{\hat{p}_5=0} \\
    &= \left[\mc{M}^C_i\times\frac{\hat{p}_5+\hat{p}_6}{\hat{s}_{i56}} - q_i^2\frac{\ket{6}[5|}{\vev{5\hat{i}}[6\hat{i}]} \right]\Big|_{\hat{p}_5=0} \ .
}
where the Schouten Identity $[56][\hat{i}| = [5\hat{i}][6| + [\hat{i}6][5|$ is applied. The second term in the result cancels the $i=j$ components of the first sum in eq.~\eqref{eq:4+2o11}.
Now it is completely written in a form identical to $\hat{S}_6^{(1)}\hat{S}_5^{(1)}$, so that it turns to the residue at $z=0$. In sum, we have all the terms in eq.~\eqref{eq:4+2o} as
\eq{
    & \mc{M}^{(o)}_{4+2} = \Bigg[ S^{(0)}_5 S^{(0)}_6 + S^{(0)}_6 S^{(1)}_5 + S^{(0)}_5 S^{(1)}_6 \\
    &\hspace{2em} + \sum_{i\neq j}\frac{q_iq_j}{\vev{5i}[6j]} [5\partial_i]\vev{6\partial_i} + \sum_i \frac{\mc{M}^C_i}{s_{i56}}\left(1+(p_5+p_6)^\mu\frac{\partial}{\partial p_i^\mu}\right) \Bigg] \mc{M}_4 
}

With the building blocks $\mc{M}_{4+2}$ for both the same-helicity and opposite-helicity cases, we can in general build any kinds of $\mc{M}_{n+2}$ amplitudes at tree-level within the range of constructibility. The derivation starts with the charge decomposition, for example
\eq{
    &\mathcal{M}_{6+2}(\phi_1^{+1},\phi_2^{+1},\phi_3^{+1},\phi_4^{-1},\phi_5^{-1},\phi_6^{-1};\gamma_7,\gamma_8)\\
    &= -\mc{M}_{6+2} (\phi_1^{+1},\phi_2^{},\phi_3^{},\phi_4^{-1},\phi_5^{},\phi_6^{};\gamma_7,\gamma_8) 
    -\mc{M}_{6+2} (\phi_1^{},\phi_2^{+1},\phi_3^{},\phi_4^{},\phi_5^{-1},\phi_6^{};\gamma_7,\gamma_8) \\
    &\ -\mc{M}_{6+2} (\phi_1^{},\phi_2^{},\phi_3^{+1},\phi_4^{},\phi_5^{},\phi_6^{-1};\gamma_7,\gamma_8)
    +\mc{M}_{6+2} (\phi_1^{+1},\phi_2^{+1},\phi_3^{},\phi_4^{-1},\phi_5^{-1},\phi_6^{};\gamma_7,\gamma_8) \\
    &\ +\mc{M}_{6+2} (\phi_1^{},\phi_2^{+1},\phi_3^{+1},\phi_4^{},\phi_5^{-1},\phi_6^{-1};\gamma_7,\gamma_8) 
    +\mc{M}_{6+2} (\phi_1^{+1},\phi_2^{},\phi_3^{+1},\phi_4^{-1},\phi_5^{},\phi_6^{-1};\gamma_7,\gamma_8)\ ,
}
where each component can be computed using the gauged soft recursion relation. The method can be promoted for amplitudes with more photons, which we do not elaborate on further.

\subsection{Non-Abelian Gauge Group}

We conclude this section with a simple example for the non-Abelian case. Starting from the 4-point building block among 4 adjoint scalars
\eq{
    \mc{M}_4 = {\rm tr}[1234]\frac{s_{12}+s_{23}}{f^2}\;+\text{permutations of (2,3,4)}
}
we attempt to construct the amplitude with an additional gluon, $\mc{M}_{4+1}$. To apply the decomposition method, we first express the amplitude in terms of its fundamental color flows by extracting the generators:
\begin{align}\begin{split}
    \mathcal{M}_4(\phi_1^{A_1},\phi_2^{A_2},\phi_3^{A_3},\phi_4^{A_4})
    \equiv \left(T^{A_1}\right)^{a_1}_{b_1}\left(T^{A_2}\right)^{a_2}_{b_2}\left(T^{A_3}\right)^{a_3}_{b_3}\left(T^{A_4}\right)^{a_4}_{b_4}\mathcal{M}_4(\phi_{b_1}^{a_1},\phi_{b_2}^{a_2},\phi_{b_3}^{a_3},\phi_{b_4}^{a_4})\ .
\end{split}\end{align}
Now that the amplitude can be written as a sum over different current configurations
% Extracting the generators from the amplitude, the remaining part is defined as the fundamental and anti-fundamental representations of the $SU(N)$ group.
\begin{align}\begin{split}
    \mathcal{M}_4=\sum_{\mc{C}}\mc{M}_4(\mc{C}) = \delta_{a_2}^{b_1}\delta_{a_3}^{b_2}\delta_{a_4}^{b_3}\delta_{a_1}^{b_4} \frac{s_{12}+s_{23}}{f^2}\;+\text{permutation of (2,3,4)}\ .\label{eq:nA4}
\end{split}\end{align}
Each $\delta_{a_j}^{b_i}$ represents a color flow under an independent $SU(N)_{ij}$ group with gauge coupling $g_{ij}$. For the configuration $\mc{C} \sim \delta_{a_2}^{b_1}\delta_{a_3}^{b_2}\delta_{a_4}^{b_3}\delta_{a_1}^{b_4}$, the non-Abelian charge decomposition in eq.~\eqref{eq:decom_NA} gives
\eq{
    \mathcal{M}_{4+1}(\phi_1{}_{b_1}^{a_1},&\phi_2{}_{b_2}^{a_2},\phi_3{}_{b_3}^{a_3},\phi_4{}_{b_4}^{a_4};G_5^{A_5}) = g_{12}\left(T^{A_5}\right)^{b_1}_{a_2}\delta_{a_3}^{b_2}\delta_{a_4}^{b_3}\delta_{a_1}^{b_4}\mathcal{A}_{12}(\mc{C};G_5) \\
    &+ g_{23}\delta^{b_1}_{a_2}\left(T^{A_5}\right)_{a_3}^{b_2}\delta_{a_4}^{b_3}\delta_{a_1}^{b_4}\mathcal{A}_{23}(\mc{C};G_5) + g_{34}\delta^{b_1}_{a_2}\delta_{a_3}^{b_2}\left(T^{A_5}\right)_{a_4}^{b_3}\delta_{a_1}^{b_4}\mathcal{A}_{34}(\mc{C};G_5) \\
    &+ g_{41}\delta^{b_1}{}_{a_2}\delta_{a_3}^{b_2}\delta_{a_4}^{b_3}\left(T^{A_5}\right)_{a_1}^{b_4}\mathcal{A}_{41}(\mc{C};G_5) + \text{permutations of (2,3,4)} \ ,
}
where the amplitude components $\mc{A}_{ij}(\mc{C};G_5)$ are to be bootstrapped individually.
For instance, the first component where the gluon acts on the current $\delta_{a_2}^{b_1}$ is given by the soft photon theorem as
\eq{
    \mathcal{A}_{12}(\mc{C};G_5^{(+)}) &= \left(\frac{\vev{12}}{\vev{15}\vev{52}}+\frac{[5\partial_1]}{\vev{15}} - \frac{[5\partial_2]}{\vev{25}} \right) \frac{s_{12}+s_{23}}{f^2}= \frac{1}{f^2}\frac{\vev{13}\vev{24}[34]}{\langle 15\rangle\langle 25\rangle}\ .
}
The other components are computed similarly. Finally, identifying all couplings $g_{ij}=g$ and reassembling the generators into the traces, we obtain the full amplitude as
\eq{
    \mc{M}_{4+1}(\phi_1^{A_1},\phi_2^{A_2},\phi_3^{A_3},\phi_4^{A_4};G_5^{A_5,(+)}) = \tr[12345]\frac{g}{f^2}\frac{\langle 13\rangle[32]\langle 24\rangle}{\langle 15\rangle\langle 45\rangle} + \ \text{permutations of (1,2,3,4).}
}
This simple example serves as a proof of concept for the full non-Abelian generalization of our formalism. It illustrates the core procedure: decomposing an amplitude into gauge-invariant components defined by specific color flows, bootstrapping these simpler components using on-shell methods like the soft theorem, and finally combining them to reconstruct the full amplitude. The method successfully derives the interaction between gluons and Goldstone bosons solely from the on-shell data of the pure scalar sector and the universal soft limits.

\section{Conclusion and Discussion}

We have successfully derived and demonstrated a new on-shell recursion relation for tree-level amplitudes in gauged non-linear sigma models. The primary obstacle -- the breaking of Adler’s zero for charged scalars by gauge interactions -- has been overcome through a synthesis of new ideas: leveraging soft photon/gluon theorems to control large-$z$ behavior, a novel charge decomposition into gauge-invariant subamplitudes where Adler’s zero is restored for neutral legs, and a prescription for internal gauge bosons based on angular momentum selection rules.

This \textbf{gauged soft recursion} formalism provides a systematic and efficient framework for constructing amplitudes with arbitrary multiplicities of Nambu-Goldstone bosons and gauge bosons, in both Abelian and non-Abelian settings. The on-shell constructibility was analyzed within the $(\rho,\sigma)$ classification of scalar EFTs, showing that amplitudes with a single gauge boson are constructible if $\rho<\sigma$, while those with an arbitrary number require $\rho<1$, a condition satisfied by the NLSM, the focus of our explicit examples.

Historically, the poor large-momentum behavior of effective field theories was a major barrier to on-shell constructibility. This limitation was first overcome for purely scalar theories by soft recursion relations, which introduced the Adler's zero -- a consequence of non-linearly realized symmetries -- as a new first principle for the on-shell bootstrap. In this work, we have introduced the soft theorems for gauge bosons as another first principle, thereby extending the range of on-shell constructibility to include gauge interaction. 
The prescription for internal photon contributions, which utilizes angular momentum as a defining criterion, further exemplifies how physical constraints beyond factorization can be harnessed to resolve on-shell ambiguities.
Our work demonstrates that systematically incorporating first principles that reflect the implicit physical constraints of a theory can maximally extend the scope and power of the on-shell bootstrap.

\section*{Acknowledgements}

M.-L.X. is supported by the National Natural Science Foundation of China (Grant No.12405123), Fundamental Research Funds for the Central Universities, Sun Yat-sen University (Grant No.25hytd001), Shenzhen Science and Technology Program (JCYJ20240813150911015). I.L.  is supported in part by the U.S. Department of Energy under contracts DE-AC02-06CH11357 (Argonne), DE-SC0023522 (Northwestern), DE-SC0010143 (Northwestern) and No. 89243024CSC000002 (QuantISED Program). Y.-H.Z. is supported by a KIAS Individual Grant (PG096402) through the School of Physics at the Korea Institute for Advanced Study.

\bibliographystyle{unsrt}
\bibliography{ref}

\end{document}